\documentclass[reprint,amsmath,amssymb,aps,prx,floatfix,longbibliography]{revtex4-2}

\usepackage{graphicx, dcolumn, bm, mathtools, bbm, mathrsfs, xcolor, float, empheq, comment, upgreek, placeins}
\usepackage[scr=boondox]{mathalpha}
\usepackage[caption=false]{subfig}
\usepackage{tikz}
\usetikzlibrary{matrix,calc, arrows.meta, automata, positioning, quotes}
\usepackage{tabularray} 
\UseTblrLibrary{booktabs}

\definecolor{fireEngine}{RGB}{193,39,45}
\definecolor{navyBlue}{RGB}{0,0,139}
\definecolor{teal}{RGB}{69,126,136}
\definecolor{caledon}{RGB}{172,234,189}
\definecolor{naplesYellow}{RGB}{246,210,88}
\definecolor{forest}{RGB}{34,139,34}

\usepackage[colorlinks=true,citecolor=forest,linkcolor=fireEngine, urlcolor=navyBlue]{hyperref}

\newcommand{\paren}[1]{\left( #1 \right)}
\newcommand{\sbrac}[1]{\left[ #1 \right]}
\newcommand{\curly}[1]{\left\lbrace #1 \right\rbrace}
\newcommand{\angbr}[1]{\left\langle #1 \right\rangle}
\newcommand{\abs}[1]{\left| #1 \right|}
\newcommand{\norm}[1]{\left\| #1 \right\|}
\newcommand{\e}[1]{\mathrm{e}^{#1}}
\newcommand{\md}{\mathrm{d}}
\newcommand{\kB}{k_{\mathrm{B}}}

\newcommand{\rma}{\mathbb{W}}          
\newcommand{\eq}{\bm{\pi}}               
\newcommand{\simp}{\Delta^{n}}           
\newcommand{\one}{\mathbf{1}}            
\newcommand{\Deq}{D_{\eq}}          
\newcommand{\Wex}{\mathscr{W}}               
\newcommand{\II}{\mathrm{I\!I}}   
\newcommand{\cond}{\mathscr{c}}           
\renewcommand{\t}{^{\mathsf{T}}}

\newcommand{\Aut}{\operatorname{Aut}}
\newcommand{\Iso}{\operatorname{Iso}}

\newcommand{\Sym}{\operatorname{Sym}}        
\newcommand{\Span}{\operatorname{span}}

\newcommand{\diag}{\operatorname{diag}}

\newcommand{\tr}{\operatorname{tr}}
\newcommand{\kcur}{\mathbb{K}}
\renewcommand{\eqref}[1]{\hyperref[#1]{(\ref*{#1})}}

\begin{document}

\title{Extrinsic thermodynamic geometry, symmetry, and the cost of constraint}

\author{Jordan R.\ Sawchuk}
\email{jordan\_sawchuk@sfu.ca}
\author{David A.\ Sivak}
\affiliation{Department of Physics, Simon Fraser University, Burnaby, British Columbia, Canada V5A1S6}

\date{\today}

\begin{abstract}
    Driving a stochastic system in finite time costs excess work; exactly how much depends on which system parameters, and hence which paths in distribution space, a controller can access. In the friction-tensor formalism for slow driving, the space of equilibrium distributions is a Riemannian manifold and constraints on a controller define submanifolds. Whether a constrained controller is forced onto suboptimal paths is a question of the extrinsic thermodynamic geometry of the submanifold, which we develop for discrete systems. We show that submanifolds with zero extrinsic curvature can be identified from symmetries of the transition network, before any optimization. A minimum-work protocol on such a submanifold is therefore stationary in the excess work even when the constraints are lifted. The stationarity of symmetric optima holds exactly, beyond slow driving. In the slow-driving regime, we show that the ambient curvature transverse to a symmetric submanifold determines whether nearby symmetry-broken protocols have lower excess work. This curvature, which separates into independent contributions from each pattern of symmetry breaking, thus acquires an operational meaning as a measure of the thermodynamic incentive to break a given symmetry. A thermodynamic comparison theorem provides an inexpensive test of local minimality that weighs the extremal curvatures along the symmetric protocol against its excess work. There is no need for such tests under certain convexity conditions on the transition rates and control parameters, which guarantee that a symmetric optimum is a global optimum. These results certify the global optimality of a five-parameter protocol previously observed to match 15-parameter full control of a four-spin Ising system, and explain why partial symmetry breaking is optimal for the same system under local-field control. 
\end{abstract}

\maketitle

\section{Introduction \label{sec:intro}}
Driving a stochastic system between prescribed distributions in finite time dissipates more than doing so quasistatically. Quantifying and minimizing this excess cost, aptly dubbed the ``price of haste''~\cite{andresenCurrentTrendsFinitetime2011}, is a long-standing aim of nonequilibrium thermodynamics~\cite{salamonFiniteTimeOptimizations1981, salamon_thermodynamic_1983, sekimotoComplementarityRelationIrreversible1997, espositoFiniteTimeThermodynamics2010, muratore-ginanneschi_heat_2013, deffnerThermodynamicControlOld2020a, blaberOptimalControlStochastic2023, van_vu_thermodynamic_2023}. Implicit in a growing body of work is a related question: What is the \textit{cost of constraint}? Constraining a controller can only raise the minimum dissipation of a transformation, but by how much?

Unsurprisingly, the answer varies greatly from system to system. For driven barrier crossing in a harmonic trap, control over both the stiffness and center of the trap reduces the excess work by ${\sim}70\%$ relative to the optimized center-only protocol~\cite{blaber_efficient_2022}, while the same pair of control parameters applied to a model of driven $\mathrm{F}_1$-ATPase reduces it by only ${\sim}5\%$~\cite{warehamMultiparameterOptimalControl2025}. In a $3\!\times\!3$ Ising model driven between magnetized configurations, controlling four independent spatially resolved magnetic fields reduces dissipation by ${\sim}92\%$ relative to the optimized protocol in a single global field~\cite{louwerseMultidimensionalMinimumworkControl2022}. 

The question is sharpened when the added parameters allow a controller to break a symmetry of the system,
so that the enlarged control set contains a symmetric subset on which the original control problem is recovered. One might expect the optimal protocol to inherit the symmetry of the problem~\cite{palaisPrincipleSymmetricCriticality1979}; indeed, this is commonly observed~\cite{schmiedlOptimalFinitetimeProcesses2007, acconciaDegenerateOptimalPaths2015, rotskoff_optimal_2015, zhongLimitedcontrolOptimalProtocols2022, nazeOptimalFinitetimeProcesses2022, loosUniversalSymmetryOptimal2024}, and symmetric control sets are routinely adopted~\cite{louwerseMultidimensionalMinimumworkControl2022, monterEnergyEfficientControlInteracting2026}. Nonetheless, the optimal strategy sometimes involves breaking symmetries. In a model of finite-time separation of two gases with two controllable selectively permeable membranes, for instance, the (effectively one-parameter) symmetric protocol is narrowly beaten by either of a mirror-image pair of symmetry-breaking protocols~\cite{chenGeodesicLowerBound2023}. 

The immediate motivation for this work is a recent paper by the authors~\cite{sawchuk_global_2026}. Across 200 different control sets applied to the task of reversing the magnetization in a $2\!\times\!2$ spin system, only 42 distinct optimal protocols appeared, with the optimal protocol of one particular five-parameter set identical (to numerical precision) to the 15-parameter full-control optimum. Starting from the five-parameter set, the three-parameter set obtained by revoking control of the pairwise spin couplings nearly tripled the optimal cost, while a different three-parameter set obtained by removing control of the higher-order couplings raised it by less than $0.2\%$. An added parameter may thus be redundant, marginally beneficial, or transformative. 

There exist bounds on the work when a constraint places a target distribution out of reach~\cite{kolchinskyWorkEntropyProduction2021}, though these are silent on constraints that leave the endpoints reachable while restricting the paths between them. General bounds also exist on the dissipation of any transformation between two distributions under arbitrary constraints on the transition rates~\cite{van_vu_thermodynamic_2023}, though these apply to each set of constraints separately and do not by themselves determine whether lifting a constraint lowers the minimum dissipation. Here we ask a related set of questions: When can minimally dissipative driving be achieved with only a few well-chosen parameters, can those parameters be identified before any optimization, which constraints on control are worth lifting, and which symmetries are worth breaking?

Our starting point is the \textit{friction-tensor formalism}, a geometric formulation of the excess work in a slowly driven stochastic system~\cite{sivakThermodynamicMetricsOptimal2012, zulkowskiGeometryThermodynamicControl2012}. For conservative driving of detailed-balanced systems, a generalized friction tensor endows the space of equilibrium distributions with a Riemannian metric, and a minimum-work protocol between two distributions is a minimizing geodesic, a shortest path on this thermodynamic manifold. Constraints restrict a controller to a partial-control submanifold~\cite{sawchuk_global_2026, sawchuk_thermodynamic_2026}, so the cost of constraint is the excess geodesic length between the same two endpoints when the path is confined to a submanifold. This is invisible to the submanifold's intrinsic geometry, hence the natural language is \textit{extrinsic geometry}, the study of submanifolds within larger spaces. 

Extrinsic geometry has been used to explore analogous questions in information geometry, where a statistical model is a submanifold (a curved exponential family) of a larger exponential family carrying the Fisher--Rao metric. There, a model's extrinsic curvature was shown to set the minimum information an estimator loses in projecting the data onto the model~\cite{efronDefiningCurvatureStatistical1975, amariDifferentialGeometryCurved1982}. In quantum control, counterexamples to a conjectured speed limit for state preparation were recently found by appeal to the extrinsic curvature of the accessible ground state within the larger set of states reachable by nonadiabatic driving~\cite{gollGeometricSpeedLimit2026}.

In this paper we introduce \textit{extrinsic thermodynamic geometry} as a systematic framework for studying the cost of constraint in driven stochastic systems, proving three main results. First, when a constraint respects a dynamical symmetry of a system (identified with automorphisms of its transition network), the constrained optimum remains stationary in the excess work when the constraint is lifted (an instance of Palais's principle of symmetric criticality~\cite{palaisPrincipleSymmetricCriticality1979}). Second, the curvature normal to the submanifold along an optimal symmetric protocol determines whether that protocol is a local minimum or a saddle point of the excess work on the enlarged protocol space (the protocols available once symmetry constraints are lifted). This curvature is furthermore block-diagonalized in a symmetry-adapted basis derived from the transition network alone, with blocks corresponding to orthogonal symmetry-breaking sectors. Each is a distinct refinement of the control set, and whether any combination of such refinements reduces the excess work may be tested one at a time. Stationarity and the sector decomposition rely only on the invariance of the excess work under automorphisms and so hold at arbitrary driving speeds, though the curvature test and geometric interpretations are specific to linear response. A \textit{thermodynamic comparison theorem}, which requires only the extremal curvatures along a symmetric protocol, sets a threshold dissipation above which breaking a symmetry is guaranteed to lower the cost. Third, for kinetics with concave edge conductances (e.g., Glauber dynamics), the excess work is a convex functional (as is the analogous action in discrete optimal transport~\cite{maas_gradient_2011, erbarRicciCurvatureFinite2011}) so for any control set whose protocol space is convex, symmetric optima are global optima and nothing is gained by breaking symmetries.

\section{Overview \label{sec:results}}
\subsection{Dynamics and thermodynamics \label{ssec:setup}} 
We consider a continuous-time Markov jump process on a finite state space $\Omega = \curly{0,\dots,n}$ with distribution $\bm{p}(t)$ governed by the master equation
\begin{equation}
    \dot{\bm{p}}(t) = \rma(t)\bm{p}(t) \ ,
\end{equation}
with irreducible rate matrix $\rma(t)$. The off-diagonal entries $w_{\mu\nu}(t)\!\geq\!0$ of $\rma(t)$ correspond to the transition rates $\nu\!\to\!\mu$, and the diagonal entries are fixed by conservation of probability [$\one\t \rma(t) = 0$]. The rates depend on time through a control protocol $\bm{u}(t) = (u_1(t),\dots, u_m(t))$. 

We restrict attention to \textit{conservative control}, so the parameters modulate the state energies $\bm{V}(\bm{u})$ (in units of the thermal energy $\kB T$ of the bath) and driving changes $\eq(t) \propto \e{-\bm{V}(\bm{u}(t))}$, the equilibrium distribution to which the system would relax if the parameters were held fixed at their values at time $t$. The rates remain detailed balanced with respect to this distribution at each instant,
\begin{equation} \label{eq:DB}
     w_{\mu\nu}(t)\pi_\nu(t) = w_{\nu\mu}(t)\pi_\mu(t) \ .
\end{equation}
The system begins in equilibrium $\eq_0 \equiv \eq(\bm{u}(0))$ at the initial parameters, and is then driven for a finite duration $\tau$ to prescribed final parameters $\bm{u}(\tau)$. In the quasistatic limit ($\tau\!\to\!\infty$) the mean work equals the free-energy change $\Delta F$ between $\eq_0$ and $\eq_1\!\equiv\!\eq(\bm{u}(\tau))$. In finite time, the distribution $\bm{p}(t)$ lags the equilibrium distribution $\eq(t)$ so the final distribution $\bm{p}(\tau)$ generally differs from $\eq_1$, and the mean work in excess of the quasistatic lower bound $\Delta F$ is
\begin{subequations}  \label{eq:excessWorkExact}
    \begin{align}
        \angbr{\mathcal{W}_{\rm ex}}  &\equiv \angbr{\mathcal{W}} - \Delta F \\
        &= \kB T\int_0^\tau \md t\, \paren{\bm{p}-\eq}\t \dot{\bm{V}} \ .
    \end{align}
\end{subequations}
When $\tau$ is long compared with the system's relaxation times (though still finite), the excess work takes the linear-response form~\cite{sivakThermodynamicMetricsOptimal2012, bonancaOptimalDrivingIsothermal2014}
\begin{subequations} 
    \begin{align}
        \angbr{\mathcal{W}_{\rm ex}} &\simeq \frac{\kB T}{\tau}\int_0^1 \md t\, \dot{\bm{u}}\t \zeta(\bm{u})\,\dot{\bm{u}} \, , \label{eq:excessWork1} \\
        \zeta_{ij}(\bm{u}) &\equiv \int_0^\infty \md t'\, \angbr{\delta f_i(t')\,\delta f_j(0)}_{\bm{u}} \ .\label{eq:frictionTensor}
    \end{align}
\end{subequations}
The time in \eqref{eq:excessWork1} is rescaled by $\tau$ (the time $t'$ in \eqref{eq:frictionTensor} is not scaled), and $\angbr{\cdot}_{\bm{u}}$ is the equilibrium average at fixed parameters $\bm{u}$. The \textit{friction tensor} $\zeta$ is the time-integrated equilibrium covariance of the forces $f_i(t) \equiv - \partial V_{\mu(t)}/\partial u_i$ conjugate to the control parameters, with $\mu(t)$ the state occupied at time $t$. Assuming that these parameters are nonredundant, in that there is no direction in parameter space along which the equilibrium distribution is unchanged to first order, the friction tensor is a Riemannian metric on the control space, and~\eqref{eq:excessWork1} is the energy of the curve $\bm{u}(t)$ in that metric~\cite{sivakThermodynamicMetricsOptimal2012, zulkowskiGeometryThermodynamicControl2012}.

We assume that each rate $w_{\mu\nu}$ depends on the control only through the energy difference $V_{\mu} - V_{\nu} = - \ln (\pi_\mu/\pi_\nu)$ of the two states it connects. The rates are then \textit{locally determined}, in that $w_{\mu\nu}$ is insensitive to the energy of any state other than $\mu$ and $\nu$, and \textit{gauge invariant}, in that a uniform energy shift leaves the rate matrix unchanged, $\rma(\bm{V} + c \one) = \rma(\bm{V})$. This includes Glauber rates $w_{\mu\nu} \propto \pi_\mu /(\pi_\mu + \pi_\nu)$~\cite{glauberTimeDependentStatisticsIsing1963} and square-root rates $w_{\mu\nu} \propto \sqrt{\pi_\mu/\pi_\nu}$~\cite{lieSquareRootApproximation2013}, but excludes rates with a fixed reference energy, such as Arrhenius rates, for which the minimum-work problem is ill-posed under full control without further constraints (\hyperref[disc]{Sec.~\ref*{disc}}). The equilibrium distribution fixes the energies only up to a uniform shift, so the rate matrix can be parametrized by its equilibrium distribution, $\rma(t) = \rma(\eq(t))$. 

\subsection{The thermodynamic manifold \label{ssec:manifold}}
Under the assumptions of~\hyperref[ssec:setup]{Sec.~\ref*{ssec:setup}}, a conservative controller can be regarded as steering $\eq$ directly, and any protocol $\bm{u}(t)$ traces a curve $\eq(t) = \eq(\bm{u}(t))$ on the \emph{open probability simplex} $\simp = \curly{\eq\,:\, \one\t\eq = 1, \ \pi_{\mu} > 0}$. The friction tensor then depends on $\bm{u}$ only through $\eq$, and~\eqref{eq:frictionTensor} can be written as~\cite{sawchuk_thermodynamic_2026}
\begin{equation} \label{eq:pullback}
    \zeta(\bm{u}) = \mathbb{J}\t L^+(\eq) \mathbb{J}\,,\quad \mathbb{J}\equiv \frac{\partial \eq}{\partial \bm{u}} \, ,
\end{equation}
for Moore--Penrose pseudoinverse $L^+$ of $L \equiv - \rma \Deq$, with $\Deq \equiv \diag(\eq)$. Under detailed balance~\eqref{eq:DB}, both $L^+$ and $L$ are symmetric matrices. If distinct parameter values give distinct accessible distributions, then the control parameters serve as coordinates on a submanifold of $\simp$~\footnote{Formally, we require that $\bm{u}\mapsto \eq(\bm{u})$ is a smooth embedding: a homeomorphism onto its image with full-rank Jacobian $\mathbb{J}=\partial \eq/\partial \bm{u}$.}, and the linear-response excess work is a functional on the simplex,
\begin{equation} \label{eq:excessWork}
    \angbr{\mathcal{W}_{\rm ex}}[\eq] \simeq \frac{\kB T}{\tau}\Wex[\eq] \ , \quad \Wex[\eq] \equiv \int_0^1 \md t\, \dot{\eq}\t L^+(\eq)\,\dot{\eq} \ .
\end{equation}
Since $\kB T$ and $\tau$ enter only as multiplicative constants, we henceforth work with the scaled excess work $\Wex$, hereafter simply called the excess work. 

Viewed geometrically, $(\simp, L^+)$ is the full-control thermodynamic manifold (hereafter abbreviated to $\simp$) with friction metric $L^+$. The (scaled) excess power $\| \dot{\eq}\|^2 \equiv \dot{\eq}\t L^+\dot{\eq}$ is the squared norm of a vector on this manifold, and the (scaled) excess work $\Wex[\eq]$ is the geometric energy of a curve $\eq$. A minimum-work protocol minimizes both the energy and the thermodynamic length $\ell[\eq] = \int_0^1 \md t \, \| \dot{\eq} \|$, and is thus a minimizing geodesic between $\eq_0$ and $\eq_1$.

The matrix $L$ and its pseudoinverse $L^+$ deserve interpretation. As a quadratic form on tangent vectors of $\simp$, $L^+$ is equivalent to $-\Deq^{-1}\rma^{\mathcal{D}}$ for Drazin inverse $\rma^{\mathcal{D}}$ of the rate matrix, which characterizes the relaxation modes and timescales~\cite{mandalAnalysisSlowTransitions2016, crooksDrazinInverseRate2018}. This is proportional to the Fisher--Rao metric $\Deq^{-1}$~\cite{ito_stochastic_2018, pistoneInformationGeometryProbability2019} when the system has a single relaxation time, and otherwise weights each mode by its own relaxation time~\cite{sawchuk_global_2026, sivakThermodynamicMetricsOptimal2012}. We return to this connection in \hyperref[ssec:RaoIllust]{Sec.~\ref*{ssec:RaoIllust}}.

\begin{figure} 
    \centering
    \includegraphics[width=\linewidth]{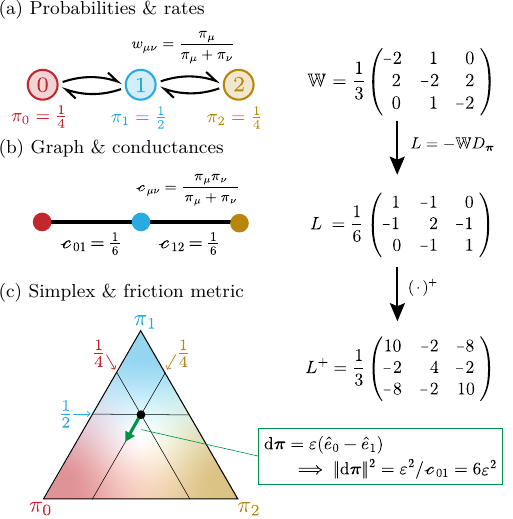}
    \caption{Three views of a three-state system with Glauber rates at $\eq=(\tfrac{1}{4},\tfrac{1}{2},\tfrac{1}{4})\t$. 
    (a)~Markov chain with rate matrix $\rma$ having entries $w_{\mu\nu}$.
    (b)~Graph with edge weights (conductances) $\cond_{\mu\nu} = w_{\mu\nu}\pi_\nu$ and weighted graph Laplacian $L = -\rma \Deq$. 
    (c)~Thermodynamic manifold $(\Delta^2,L^+)$, with the friction metric $L^+$ measuring the length $\|\md \eq\|$ of a displacement $\md \eq$ (green arrow), shown for a transfer of probability from state $1$ to state $0$.}
    \label{fig:correspondence}
\end{figure}

Complementary to its geometric interpretation, the excess work~\eqref{eq:excessWork} has a graph-theoretic interpretation that underpins many of our key results and proofs (\hyperref[fig:correspondence]{Fig.~\ref*{fig:correspondence}}). The \textit{Markov graph} $G = (\Omega, \mathcal{E})$ is defined by identifying the states of a system with the nodes $\mu \in \Omega$ of a graph, and pairs of states joined by an allowed transition ($w_{\mu\nu} > 0$, and hence $w_{\nu\mu} >0$) with its edges $(\mu,\nu) \in \mathcal{E}$. Then with edge weights $\cond_{\mu\nu} \equiv w_{\mu\nu}\pi_\nu = \cond_{\nu\mu}$ equal to the one-way equilibrium fluxes, $L$ is the \textit{weighted graph Laplacian} of $G$, i.e., the off-diagonal elements are $L_{\mu\nu} = -\cond_{\mu\nu}$ if $(\mu,\nu)$ is an edge and zero otherwise, and the diagonal elements are $L_{\mu\mu} = \sum_\nu \cond_{\nu\mu}$ (so $\one\t L = L\one = 0$)~\cite{grady_discrete_2010}. The role of $L$ as a Laplacian is visible in the master equation $\dot{\bm{p}} = -L \bm{\varrho}$ with $\varrho_{\mu}\equiv p_\mu/\pi_\mu$, the discrete analog of the Fokker-Planck equation $\partial_t p = \nabla \cdot [D\pi \nabla(p/\pi)]$ with $-L$ playing the role of the weighted continuous Laplacian $\nabla \cdot D \pi \nabla$ for diffusion coefficient $D$. The rationale for this name is also apparent for the path graph of \hyperref[fig:correspondence]{Fig.~\ref*{fig:correspondence}(b)}, where $L$ is proportional to a second-difference operator at the chosen point $\eq\in \Delta^2$. Regarding $\cond_{\mu\nu}$ as the \textit{conductance} of the edge $(\mu,\nu)$, $L^+$ determines the effective resistances of an electrical network on $G$, and the integrand of $\Wex$ is its dissipation for currents $\dot{\eq}$ injected at the nodes~\cite{sawchuk_thermodynamic_2026}.

\subsection{Submanifolds and the cost of constraint \label{ssec:constraints}}
Full control requires $n$ parameters to reach every distribution in $\simp$, but the number of controllable parameters $m$ in typical experimental settings is much smaller, and the accessible equilibria $\eq(\bm{u})$ form an $m$-dimensional submanifold $M \subseteq \simp$ with the inherited metric~\eqref{eq:pullback}. Lifting some of the constraints that define a control set enlarges $M$ to a submanifold $M'$, with $M \subset M' \subseteq \simp$. For a control task specified by endpoints $\eq_0,\eq_1 \in M$ and a controller constrained to $M$, define the \textit{protocol space} 
\begin{equation} \label{eq:pathSpace}
    \mathcal{P}_M \equiv \curly{\eq: [0,1]\to M \, \big\vert \, \eq(0)=\eq_0, \, \eq(1)=\eq_1} \ ,
\end{equation}
the set of smooth parametrized curves in $M$ between the endpoints. We will speak of $\mathcal{P}_{M}$ as a space in its own right, in which each protocol $\eq$ is a single point, and its tangent space $T_{\eq}\mathcal{P}_M$ consists of the smooth perturbations $X$ of $\eq$ available to the restricted controller [so each $X$ is a tangent field to $M$ along $\eq$ with $X(0)=X(1)=0$]. Just as lifting constraints enlarges the control submanifold, $M\subset M' \subseteq \simp$, it also enlarges the protocol space, $\mathcal{P}_M \subset \mathcal{P}_{M'} \subseteq \mathcal{P}_{\simp}$. Then the excess work~\eqref{eq:excessWork} is a function on the protocol space, $\Wex :\mathcal{P}_{\simp} \to \mathbb{R}$, geodesics are its stationary points, and minimum-work protocols are its minima (we will soon be interested in determining whether a given stationary point of $\Wex$ is a minimum or a saddle point). 

For a controller constrained to $M$, denote by $\eq^\star_M \in \mathcal{P}_M$ a minimum-work protocol, with total cost
\begin{equation}\label{eq:minWorkM}
    \Wex^\star_M \equiv \Wex[\eq^\star_M] = \min_{\eq\in\mathcal{P}_M} \Wex[\eq] \ .   
\end{equation}
Where the submanifold in question is implied or irrelevant, we drop the subscript $M$ to avoid notational clutter. 

We can now define the cost of constraint (in linear response):
\begin{equation} \label{eq:costOfConstraint}
    \Delta \Wex(M\! \to \!M') \equiv \Wex^\star_M - \Wex^\star_{M'} \geq  0 \ ,
\end{equation}
the additional excess work incurred by operating with a smaller control set. It is nonnegative because $\mathcal{P}_M\subset \mathcal{P}_{M'}$, i.e., every protocol available in $M$ is available in $M'$, and it vanishes exactly when the optimal protocol of $M'$ already lies in the smaller space $M$, in which case lifting the constraints offers no energetic advantage.

\begin{table*}[t!] 
    \caption{Important symbols and their mathematical and physical interpretations.}
    \label{tab:dictionary}
    \begin{tblr}{
    width = \textwidth,
    colspec = {Q[l,m] X[l,2.3,m] X[l,3,m]},
    row{1} = {bg=gray!12},
    row{3} = {bg=gray!8},
    row{5} = {bg=gray!8},
    row{7} = {bg=gray!8},
    row{9} = {bg=gray!8},
    row{11} = {bg=gray!8},
    rowsep = 2.5pt,
    hline{1,Z} = {0.8pt},
    hline{2} = {0.4pt},
    } 
         OBJECT(S) & MATHEMATICAL DESCRIPTION & PHYSICAL INTERPRETATION  
         \\
         $\simp$
         & $n$-dimensional open probability simplex 
         & equilibria reachable by full conservative control of an $(n+1)$-state system
         \\
         $M \subset M' \subseteq \simp$
         & nested submanifolds of $\simp$
         & equilibria reachable by a constrained controller ($M$) and those reachable after lifting some or all constraints ($M'$) 
         \\
         $L^+$ 
         & Riemannian metric on $\simp$ equal to the pseudoinverse of the weighted Laplacian $L$ of the Markov graph $G$
         & parametrization-independent friction tensor, giving the instantaneous dissipative cost of changing the equilibrium distribution
         \\
         $T_{\eq}M$; $N_{\eq}M$
         & tangent space of $M$ at $\eq$; normal space of $M$ in $M'$ ($T_{\eq}M'=T_{\eq}M\oplus N_{\eq} M$)
         & perturbations of $\eq$ available to a controller confined to $M$; perturbations available on $M'$ but not $M$
         \\
         $\mathcal{P}_{M}$; $T_{\eq}\mathcal{P}_M$ 
         & space of fixed-endpoint paths on $M$; proper variation fields along $\eq\in\mathcal{P}_M$ 
         & protocols between fixed endpoints realizable with the control parameters spanning $M$; protocol perturbations available to that controller
         \\
         $\Wex[\eq]:\mathcal{P}_{\simp} \to \mathbb{R}$  
         & geometric energy of a path $\eq \in \mathcal{P}_{\simp}$  
         & (scaled) excess work of a control protocol 
         \\
         $\eq^\star_M$; $\Wex^\star_M$ 
         & minimizing geodesic in $\mathcal{P}_M$; its energy
         & minimum-work protocol on $M$; its excess work
         \\
         $\II(\dot{\eq}^\star_M,\dot{\eq}^{\star}_M)$  
         & acceleration of a geodesic in $M$ as measured in $M'$
         & gradient of the excess work in directions made available by lifting constraints ($M\to M'$)
         \\
         $P_{\sigma}$, $\sigma \in \Aut(G)$ 
         & permutation matrix of a Markov-graph automorphism, hence an isometry of $\simp$ 
         & dynamical symmetry, a relabeling of the states that carries solutions of the master equation to solutions
         \\
         $M_H$; $\Pi_0$ 
         & fixed-point set of $H$ in $\simp$; orthogonal projector onto $U_0 =T_{\eq}M_H \oplus \Span\curly{\one}$
         & distributions assigning equal probability to states related by the symmetries in $H$; symmetrization of a distribution by averaging over $H$ 
         \\
         $\mathcal{U}_{\rho}$; $\kcur_{\rho}(t)$ 
         & orthogonal subspaces of $T_{\eq}M'$ carrying the real irreps $\rho$ of $H$; curvature operator of $M'$ restricted to $\mathcal{U}_{\rho}$ along $\dot{\eq}^\star_M$
         & inequivalent patterns of symmetry breaking available on $M'$; curvature contribution to the second-order change in excess work under perturbations in $\mathcal{U}_{\rho}$
    \end{tblr}
\end{table*}

Frequently used symbols and their physical and geometric/group theoretic interpretations are summarized in \hyperref[tab:dictionary]{Table~\ref*{tab:dictionary}}, and formal definitions from Riemannian geometry are compiled for reference in \hyperref[app:geom]{App.~\ref*{app:geom}}, though the presentation in the main text is self-contained.

\subsection{Main results \label{ssec:mainResults}}
\begin{figure*}
    \centering
    \includegraphics[width=\linewidth]{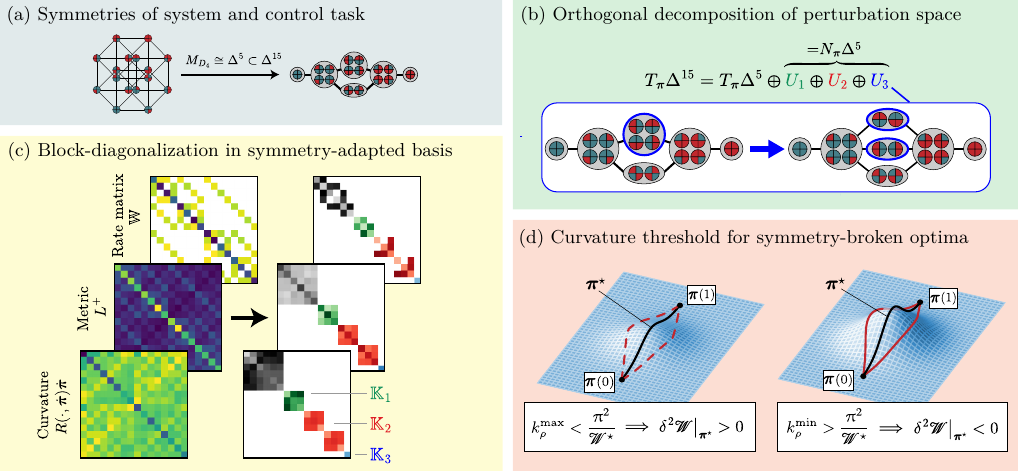}
    \caption{
    (a) The 16-state Markov graph and the lumped chain for the four-spin system and task of~\hyperref[ssec:fourSpinSym]{Sec.~\ref*{ssec:fourSpinSym}} under the symmetry group $D_4$ (graphs enlarged in \hyperref[fig:fourSpinSymmetries]{Fig.~\ref*{fig:fourSpinSymmetries}}). Every protocol on the submanifold of orbit-uniform distributions $M_{D_4}$ has the same excess work as the corresponding protocol on the thermodynamic manifold $\Delta^5$ of the lumped chain.
    (b) The directions normal to $M_{D_4}$ split orthogonally into sectors $U_1$, $U_2$, and $U_3$ (arbitrary order), each breaking the $D_4$ symmetry in a distinct way. Perturbations in $U_3$ split four configurations that are indistinguishable in $M_{D_4}$ (mesostates with thick blue outlines in the six- and seven-state lumped chains).
    (c) In a basis adapted to $D_4$, the rate matrix $\rma$, the metric tensor $L^+$, and the curvature $R(\cdot,\dot{\eq})\dot{\eq}$ (computed for randomly chosen $\eq \in M_{D_4}$ and $\dot{\eq} \in T_{\eq}M_{D_4}$) are block-diagonalized. The gray block acts within $M_{D_4}$, and each remaining block acts within one symmetry-breaking sector. In all three matrices, the two red blocks are identical. The labeled blocks $\kcur_1$, $\kcur_2$, and $\kcur_3$ are the sector curvature operators.
    (d) The thermodynamic comparison theorem tests the minimality of a symmetric minimum-work protocol $\eq^\star$ (black) with respect to small perturbations in each sector. The surfaces sketch different subspaces of a thermodynamic manifold, both with positive curvature that differs only in its size relative to the threshold $\uppi^2/\Wex^\star$. 
    For curvature below the threshold (left), every nearby symmetry-broken protocol (dashed red curves) is more dissipative; above it (right), some symmetry-broken protocols dissipate less (solid red curves). Negative curvature, not shown, spreads neighboring protocols apart, so a sector whose curvature is nowhere positive along the protocol never favors symmetry breaking whatever the cost of the task.
    }
    \label{fig:master}
\end{figure*}

In this section we give a high-level summary of our primary results, to be developed in detail in the subsequent sections and appendices. In brief, the first variation of the excess work vanishes for an optimal protocol in a symmetric submanifold; the second variation of the excess work, hence local optimality against symmetry breaking, is determined by curvature, with simple conditions on curvature and dissipation diagnosing nonoptimality; and convexity conditions on the kinetics and protocol space ensure global optimality.

\paragraph{Symmetric optima are stationary in the excess work.} 
Call a permutation $\sigma$ of the state space a \textit{dynamical symmetry} if it preserves the adjacency of the Markov graph $G$ and sends edges only to other edges hosting the same rate law, in that the rates are identical functions of the respective adjacent state probabilities. These form the automorphism group $\Aut(G)$ of the Markov graph with edges labeled by the different types of transition in the network (\hyperref[app:colored]{App.~\ref*{app:colored}}), so $\Aut(G)$ is the ordinary automorphism group when all transitions obey the same law. A permutation $\sigma$ acts on distributions through its permutation matrix $(P_{\sigma} \eq)_{\sigma(\mu)} = \pi_\mu$, and is an \textit{isometry} of the thermodynamic manifold if $\Wex[P_{\sigma}\eq] = \Wex[\eq]$ for every protocol $\eq(t)$. Our first main result is that every dynamical symmetry is a symmetry of the thermodynamic manifold,
\begin{equation} \label{eq:R1}
    \Aut(G) \subseteq \Iso(\simp) \, ,
\end{equation}
where $\Iso(\simp)$ is the isometry group of the thermodynamic manifold~(\hyperref[app:geom:iso]{App.~\ref*{app:geom:iso}}). Consequently, the fixed-point set $M_H$ of any subgroup $H \subseteq \Aut(G)$, comprising the distributions assigning equal probability to all states in each $H$-orbit, is a totally geodesic submanifold of $\simp$, meaning that every geodesic of $M_H$ is also a geodesic of $\simp$. A minimum-work protocol $\eq^\star_{M_H} \in \mathcal{P}_{M_H}$ is thus a stationary point of $\Wex$ on all of $\mathcal{P}_{\simp}$. This stationarity is exact rather than a feature of slow driving, since the excess work is invariant under $H$ at any duration. Because $\Aut(G)$ is determined by the transition network alone, control sets spanning totally geodesic submanifolds $M_H$ can be identified combinatorially, before any optimization. Optimal symmetric control on $M_H$ is furthermore equivalent to optimal full control of the $r$-state chain obtained by lumping each of the $r$ $H$-orbits into mesostates [\hyperref[fig:master]{Fig.~\ref*{fig:master}(a)}]. \hyperref[sec:sym]{Section~\ref*{sec:sym}} elaborates and extends the results to partial control by a suitable restriction $\Aut(G,M')\subseteq \Aut(G)$ for arbitrary submanifolds $M' \subseteq \simp$.

\paragraph{Curvature decides the local character of symmetric optima.} 
Let $M' \subseteq \simp$ be the submanifold accessible by the controller, and let $H$ be a group of dynamical symmetries that map $M'$ onto itself. Denote by $M$ the symmetric distributions available to the controller. By the first result, a minimum-work protocol $\eq^\star_M\in\mathcal{P}_{M}$ is a stationary point of the excess work on $\mathcal{P}_{M'}$, but it need not be a minimum. Whether breaking a symmetry can lower the excess work is then decided by a second-variational test. Our second main result is that the symmetry-breaking directions along the symmetric geodesic $\eq^\star_M$ split into $L^+$-orthogonal \textit{symmetry-breaking sectors} $\mathcal{U}_{\rho}$, one for each nontrivial real irreducible representation $\rho$ of $H$ [\hyperref[fig:master]{Fig.~\ref*{fig:master}(b)}], and the second variation of the excess work decouples as
\begin{align}
    &\Wex[\eq^\star_M + \epsilon X] - \Wex^\star_M \nonumber \\
    &\quad= \epsilon^2\,\sum_{\rho}\int_0^1 \md t \, \sbrac{|\dot{\bm{x}}_\rho|^2 - \Wex^\star_M\, \bm{x}_\rho\t\kcur_{\rho}(t)\bm{x}_{\rho}} + \cdots \    \label{eq:R2index}
\end{align}
where $X\in T_{\eq^\star_M}\mathcal{P}_{M'}$ is a perturbation with components $\bm{x}_{\rho}(t)$ in $\mathcal{U}_{\rho}$ (in an orthonormal basis of each sector carried along the protocol), and $\kcur_{\rho}(t)$ is the \textit{sector curvature operator} [labeled blocks in \hyperref[fig:master]{Fig.~\ref*{fig:master}(c)}]. Its diagonal entries are the sectional curvatures of $M'$ along $\dot{\eq}^\star_M(t)$ and each basis direction of $\mathcal{U}_{\rho}$ (\hyperref[app:geom:curvature]{App.~\ref*{app:geom:curvature}}). The protocol is a local minimum of $\Wex$ on $\mathcal{P}_{M'}$ if the $\epsilon^2$ term is positive definite and a saddle point if it is indefinite (a saddle point of the function $\Wex$ on the protocol space, unrelated to saddle-shaped regions of the thermodynamic manifold). The first term in the integrand of~\eqref{eq:R2index} is never negative, so a saddle point requires positive curvature in some sector, large compared with $1/\Wex^\star_M$. Curvature thus measures the thermodynamic incentive to break a dynamical symmetry, weighted by the cost of the task. Like stationarity, the decoupling into sectors holds at any driving speed, since it rests only on the symmetry of the excess work, but the geometric interpretation is specific to slow driving. 

Determining the sign of each term in~\hyperref[eq:R2index]{Eq.~(\ref*{eq:R2index})} generally cannot be done in closed form. Bounding $\kcur_{\rho}$ by its extreme eigenvalues $k_\rho^{\min}$ and $k_{\rho}^{\max}$ along the protocol [\hyperref[eq:kminmax]{Eq.~(\ref*{eq:kminmax})}] and comparing with spaces of constant curvature where the answer is known (hyperspheres, Euclidean space, or hyperbolic space for positive, zero, or negative curvatures) gives a special case of the Rauch comparison theorem, which we call the \textit{thermodynamic comparison theorem}. It tests each sector against a threshold $\uppi^2/\Wex^\star_M$ ($\uppi$ denoting the number pi, distinct from the equilibrium distributions $\eq$) set by the symmetric optimum's cost [\hyperref[fig:master]{Fig.~\ref*{fig:master}(d)}]. Curvature below the threshold in every sector makes $\eq^\star_M$ a local minimum of $\Wex$, and curvature above the threshold in any sector makes it a saddle point [\hyperref[eq:R2thresholds]{Eq.~(\ref*{eq:R2thresholds})}]. Between these bounds, the test is inconclusive. \hyperref[sec:comparison]{Section~\ref*{sec:comparison}} develops the theorem and demonstrates both decisive and inconclusive cases, and \hyperref[disc]{Sec.~\ref*{disc}} discusses ways to strengthen it. 

\paragraph{Symmetric optima are globally optimal when the conductance is concave and the protocol space is convex.} 
While the previous results concern local optimality, our third result is global. If the conductance $\cond_{\mu\nu}$ on every edge is jointly concave in the adjacent equilibrium probabilities $\pi_{\mu}, \pi_{\nu}$, 
then
symmetrizing a protocol never raises its cost,
\begin{equation} \label{eq:R3}
    \Wex \sbrac{\Pi_0 \eq} \, \leq \, \Wex[\eq] \ ,    
\end{equation}
provided that the controller can realize the symmetrized protocol. Here $H \subseteq \Aut(G)$ is a subgroup fixing the protocol's endpoints, and $\Pi_0$ is the \textit{orbit average}, $(\Pi_0\eq)_\mu = \frac{1}{|\mathcal{O}|}\sum_{\nu \in \mathcal{O}} \pi_\nu$ for $\mu$ in the $H$-orbit $\mathcal{O}$. Under full control, $\Pi_0 \eq \in \mathcal{P}_{\simp}$ for every $\eq \in \mathcal{P}_{\simp}$, so symmetric optima are global optima among all conservative protocols. This result establishes that the five-parameter protocol on the four-spin system of Ref.~\cite{sawchuk_global_2026} is globally optimal. If a constrained controller cannot realize all symmetrized protocols (when the protocol space is nonconvex), breaking a symmetry can still outperform the best symmetric protocol available to the controller even when the conductance is concave, as for local-field control of four spins (\hyperref[ssec:fourspin]{Sec.~\ref*{ssec:fourspin}}). \hyperref[sec:convexity]{Section~\ref*{sec:convexity}} proves the result and physically interprets the concavity condition on the edge conductances.

\section{Dynamical symmetries and totally geodesic control sets \label{sec:sym}}
In this section, we demonstrate how a minimum-work protocol confined to a symmetric submanifold is also stationary in the excess work over the full control space, and we extend this result to partial control. The formal generalization to multiple rate laws in a network, the proofs, and the coarse-graining construction are collected in \hyperref[app:symmetry]{App.~\ref*{app:symmetry}}.  

\subsection{Second fundamental form and stationarity}
We begin by introducing the basic object of extrinsic differential geometry, the second fundamental form. Let $M \subset \simp$ be a partial-control thermodynamic submanifold, and let $\eq\in M$ be a distribution. The perturbations of $\eq$ accessible to a complete controller form the tangent space $T_{\eq}\simp = \one^\perp$ of the simplex, the normalization-conserving (hence zero-sum) vectors, and the perturbations accessible to a controller restricted to $M$ form the tangent space $T_{\eq}M$ of the submanifold at $\eq$. Any perturbation $\delta \eq \in \one^{\perp}$ decomposes uniquely into an accessible part in $T_{\eq}M$ and an orthogonal, forbidden part in the normal space $N_{\eq}M$ of $M$ in $\simp$. That is, $\one^{\perp} = T_{\eq}M \oplus N_{\eq} M$, with $\oplus$ the direct sum. Along a protocol, this construction applies at each time $t$, with $\eq(t)\in M$ and $\dot{\eq}(t) \in T_{\eq(t)}M$. 

The \textit{second fundamental form} of $M$ in $\simp$ is a symmetric bilinear map $\II:T_{\eq}M\times T_{\eq}M \to N_{\eq}M$ that captures the way $M$ sits in $\simp$ [\hyperref[eq:gauss]{Eq.~(\ref*{eq:gauss})}]. In particular, $\II(\dot{\eq}(t),\dot{\eq}(t))$ is the normal component of the ambient covariant acceleration of a curve $\eq \in \mathcal{P}_M$ at time $t$. A geodesic has no tangential acceleration, so $\II(\dot{\eq}^\star_M(t), \dot{\eq}^\star_M(t))$ is its entire ambient acceleration and thereby characterizes how a path that is `straight' in $M$ bends in $\simp$. Thermodynamically, 
\begin{equation}\label{eq:IIgrad}
    \II\big(\dot{\eq}^\star_M(t),\dot{\eq}^\star_M(t)\big) =-\tfrac{1}{2}L(\eq^\star_M(t)) \, \frac{\delta \Wex}{\delta \eq(t)}\bigg\vert_{\eq^\star_M} 
\end{equation}
is the contravariant gradient of the excess work in the forbidden directions (\hyperref[app:geom:coords]{App.~\ref*{app:geom:coords}}), and $2L^+ \II(\dot{\eq}^\star_M, \dot{\eq}^\star_M)$ is the generalized force conjugate to perturbations off $M$, so a perturbation $X$ off $M$ changes the excess work to first order by $-2 \int_0^1\md t\,  X\t L^+ \II(\dot{\eq}^\star_M,\dot{\eq}^\star_M)$. (This carries over with minimal modification if the ambient space is some other less-constrained partial-control submanifold $M' \supset M$ rather than $\simp$.) 

If $\II(\dot{\eq}^\star_M,\dot{\eq}^\star_M) \neq0$ somewhere along the protocol, there are perturbations of the protocol off $M$ which decrease the excess work at first order, so lifting the constraint lowers the cost. If $\II=0$ 
identically on $M$, then the submanifold is called \textit{totally geodesic}: every geodesic of the submanifold is also a geodesic of the ambient space. Total geodesity generalizes the sense in which a plane is `flat' in three-dimensional Euclidean space while a cylinder is not, though both the plane and cylinder are intrinsically flat. By \hyperref[eq:IIgrad]{Eq.~(\ref*{eq:IIgrad})}, every minimum-work protocol of a totally geodesic submanifold is stationary in $\Wex$ under perturbations off the submanifold. The practical significance of \hyperref[eq:R1]{Eq.~(\ref*{eq:R1})} is that such submanifolds can be found directly from the graph automorphisms.

\subsection{Dynamical symmetries and totally geodesic control sets}
An automorphism $\sigma \in \Aut(G)$ carries each edge $(\mu,\nu)$ to an edge $(\sigma(\mu),\sigma(\nu))$ hosting the same rate law, so the rate matrix is equivariant under the permutation $\sigma$: $\rma(P_{\sigma}\eq) = P_{\sigma} \rma(\eq)P_{\sigma}\t$ (\hyperref[app:colored]{App.~\ref*{app:colored}}). Automorphisms of the Markov graph are dynamical symmetries in the sense that if $\bm{p}(t)$ solves the master equation under the protocol $\eq(t)$, then $P_{\sigma}\bm{p}(t)$ solves it under $P_{\sigma}\eq(t)$. This equivariance of the rates carries over to the Laplacian $L = -\rma \Deq$:
\begin{equation} \label{eq:isocriterion}
    L(P_{\sigma}\eq) = P_{\sigma}L(\eq)P_{\sigma}\t \ , \quad \forall \,\eq \in \simp \, .
\end{equation}
The metric $L^+$ inherits this equivariance (\hyperref[app:symproofs]{App.~\ref*{app:symproofs}}), from which it follows that $\Wex[P_{\sigma} \eq] = \Wex[\eq]$ and thus $P_{\sigma}$ is an isometry of the thermodynamic manifold. Conversely, every permutation acting isometrically on the thermodynamic manifold is a dynamical symmetry.

\begin{figure}
    \centering
    \includegraphics[width=\linewidth]{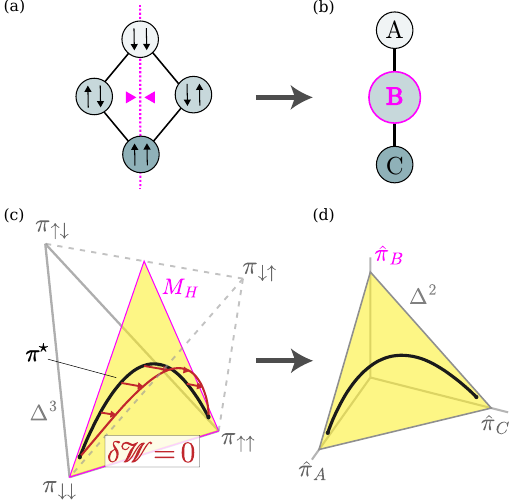}
    \caption{The Markov graph of a two-spin Ising system (a) is lumped under a spin-exchange symmetry to a three-state chain (b) whose states $A$, $B$, and $C$ are respectively the orbits $\curly{\downarrow\downarrow}$, $\curly{\uparrow\downarrow, \downarrow\uparrow}$, and $\curly{\uparrow\uparrow}$. The submanifold $M_H$ of distributions with $\pi_{\uparrow\downarrow} = \pi_{\downarrow\uparrow}$ (c) has the same thermodynamic geometry as $\Delta^2$ for the lumped chain (d), and the optimal protocol $\eq^\star$ constrained to $M_H$ has excess work stationary to perturbations in $T_{\eq^\star_{M_H}}\mathcal{P}_{\Delta^3}$. Dark red arrows indicate a protocol perturbation off $M_H$. Magenta marks the exchange symmetry in (a) and the objects it defines, namely the mesostate $B$ in (b), the submanifold $M_H$ in (c), and the orbit mass $\hat{\pi}_B$ in (d).}
    \label{fig:CG2spin}
\end{figure}

By a standard theorem of Riemannian geometry, the fixed-point set of any family of isometries of a manifold is a totally geodesic submanifold~\cite[Thm.~II.5.1]{kobayashi_geomtry}. Since every dynamical symmetry is an isometry, the distributions left unchanged by every element of a subgroup $H \subseteq \Aut(G)$ thus form a totally geodesic submanifold 
\begin{equation} \label{eq:MH}
    M_H \equiv \curly{\eq \in \simp \,: \, \pi_{\sigma(\mu)}=\pi_\mu, \ \forall \mu\in \Omega, \, \forall \sigma \in H} \ .
\end{equation}
This comprises the distributions uniform on each of the $r$ orbits of $H$, hence $\dim M_H = r\!-\!1$ by normalization. 

We may view this from a coarse-graining perspective, defining for each orbit $\mathcal{O}$ the \textit{orbit mass} $\hat{\pi}_{\mathcal{O}} = \sum_{\mu \in \mathcal{O}}\pi_\mu$. Since $\eq$ and $\rma(\eq)$ are $H$-invariant on $M_H$, the dynamics do not distinguish the states within an orbit, and the orbit masses evolve via a Markov process of their own, the \textit{lumped chain}~\cite{kemenyFiniteMarkovChains1983, tianLumpabilityCommutativityMarkov2006}. The conductance between two states of the lumped chain is the sum of the conductances of all edges joining the corresponding orbits on the original graph, and these conductances define the weighted graph Laplacian $\hat{L}$ of the lumped chain. Each distribution on $M_H$ is fixed by its orbit masses, and a protocol on $M_H$ has the same excess work as the protocol $\hat{\eq}(t)$ on the thermodynamic manifold $(\Delta^{r-1}, \hat{L}^+)$ of the lumped chain (\hyperref[app:CG]{App.~\ref*{app:CG}}), so optimal symmetric control of the original system is equivalent to optimal full control of the coarse-grained system,
\begin{equation}\label{eq:lumpedCost}
    \Wex^\star_{M_H} = \Wex^\star_{\Delta^{r-1}} \ .
\end{equation}
Under the concavity conditions of \hyperref[sec:convexity]{Sec.~\ref*{sec:convexity}}, the right-hand side is also the optimal cost of full control of the original system. In this case it is the number of orbits, rather than the number of states, that sets the dimension of the full-control problem.

\begin{figure}
    \centering
    \includegraphics[width=\linewidth]{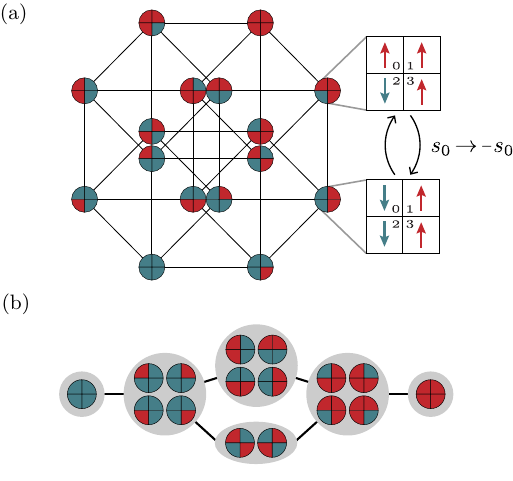}
    \caption{(a) The Markov graph of a four-spin system is isomorphic to a tesseract (4-cube), adapted from Ref.~\cite{sawchuk_global_2026}. (b) Lumped chain from the subgroup $D_4$ of symmetries of the square.}
    \label{fig:fourSpinSymmetries}
\end{figure}

Two spins with equal bare flip rates provide a minimal example (\hyperref[fig:CG2spin]{Fig.~\ref*{fig:CG2spin}}). Exchanging the antialigned states $\uparrow\downarrow$ and $\downarrow\uparrow$ is an automorphism of the four-state Markov graph, and for the group $H\cong C_2$ generated by this exchange (with $C_2$ the cyclic group of order two), $M_H$ is the plane of distributions with equal probabilities in the antialigned states. The orbits $\curly{\downarrow\downarrow}$, $\curly{\uparrow\downarrow,\downarrow\uparrow}$, and $\curly{\uparrow\uparrow}$ lump to a three-state chain whose middle state is joined to each terminal state by two parallel edges of the original graph, so symmetric control on $M_H$ is full control of the three-state chain. A more dramatic reduction occurs for the four-spin system of \hyperref[ssec:fourSpinSym]{Sec.~\ref*{ssec:fourSpinSym}}, where the 16 states lump to six orbits under $D_4$ (\hyperref[fig:fourSpinSymmetries]{Fig.~\ref*{fig:fourSpinSymmetries}}).

\subsection{Symmetries under partial control}
The lumped chain fully characterizes the control problem only when the controller can tune every orbit mass independently (up to normalization). Constraints may prohibit this, so we extend the construction to an arbitrary control submanifold $M' \subseteq \simp$. The group of dynamical symmetries that fix (stabilize) $M'$ is the \textit{stabilizer subgroup}
\begin{equation} \label{eq:stabilizerGroup}
    \Aut(G,M') \equiv \curly{\sigma \in \Aut(G) \, : \, P_\sigma M' = M'} \ .
\end{equation}
In other words, these are the permutations $P_{\sigma}$ that map accessible distributions to accessible distributions. Since each such permutation is an isometry of the ambient space that maps $M'$ onto itself, it is also an isometry of $M'$: 
\begin{equation} \label{eq:stabilizer}
    \Aut(G,M') \subseteq \Iso(M') \ .
\end{equation}
For $H \subseteq \Aut(G,M')$, the symmetric control set available to the restricted controller is the intersection $M=M_H \cap M'$, which we take to be connected. (If $M_H\cap M'$ is not connected, the statements here hold for each connected component separately). Since $M$ is the fixed-point set of a group of isometries of $M'$, it is totally geodesic in $M'$, even when $M'$ is curved in $\simp$ (\hyperref[app:symproofs]{App.~\ref*{app:symproofs}}). For endpoints $\eq_0,\eq_1 \in M$, a minimum-work protocol $\eq^\star_M \in \mathcal{P}_{M}$ for a controller constrained to $M$ is therefore also stationary in the excess work on the enlarged protocol space $\mathcal{P}_{M'}$.

While the geometric interpretation is a feature of the slow-driving regime, the stationarity result is more general. Let $H\subseteq \Aut(G,M')$ be a group of dynamical symmetries fixing both endpoints $\eq_0, \eq_1 \in M'$ of an arbitrarily fast protocol (as before, $\bm{p}(0) = \eq_0$, and $\eq_1$ is the distribution to which the system will relax from the final distribution $\bm{p}(\tau)$), so that $H$ maps $\mathcal{P}_{M'}$ onto itself. Because dynamical symmetries map solutions of the master equation onto other solutions, the exact excess work~\eqref{eq:excessWorkExact} is invariant under $H$ at any duration $\tau$. Then by the principle of symmetric criticality, which states that under mild technical conditions ``critical symmetric points are symmetric critical points''~\cite{palaisPrincipleSymmetricCriticality1979}, a protocol stationary among $H$-invariant protocols is stationary among all protocols on $\mathcal{P}_{M'}$ (\hyperref[ssec:exactExcess]{App.~\ref*{ssec:exactExcess}}). As for geodesics, this says nothing about whether such a protocol is a minimum.

\subsection{Example: Four spins \label{ssec:fourSpinSym}}
The spin system of Ref.~\cite{sawchuk_global_2026} will serve as a recurring example throughout this paper. For four Ising spins $s_i = \pm 1$, any energy function on the $2^4=16$ states (up to constant offset) can be written as a polynomial in the spins,
\begin{align} \label{eq:fourspinV}
    V(\bm{s}) = &-\sum_{i} h_i s_i - \sum_{i<j} J_{ij}\, s_i s_j \nonumber \\
    &- \sum_{i<j<k} K_{ijk}\, s_i s_j s_k - Q\, s_0 s_1 s_2 s_3 \ .
\end{align}
The coefficients are interpreted as local fields $h_i$, pairwise couplings $J_{ij}$, three-spin couplings $K_{ijk}$, and a four-spin coupling $Q$ (denoted $L$ in Ref.~\cite{sawchuk_global_2026}). These $15 \, (\,=4+6+4+1)$ parameters span the simplex $\Delta^{15}$, so are sufficient for full control of the system. 

Continuous-time Glauber dynamics with transition rates $w_{\mu\nu} = [1+\e{V_{\mu}-V_{\nu}}]^{-1} = \pi_\mu/(\pi_\mu+\pi_\nu)$ govern the evolution of the system in time. Only one spin flips at a time, so the Markov graph is a 4-cube, or tesseract [\hyperref[fig:fourSpinSymmetries]{Fig.~\ref*{fig:fourSpinSymmetries}(a)}]. Its automorphisms are the $4!\times 2^4 = 384$ ways to permute the spins and reverse any subset of them, and by \hyperref[eq:R1]{Eq.~(\ref*{eq:R1})} each is an isometry of the full thermodynamic manifold $\Delta^{15}$. In the parameters of \hyperref[eq:fourspinV]{Eq.~(\ref*{eq:fourspinV})}, permuting the spins permutes the $h_i$ among themselves, and likewise for the $J_{ij}$ and for the $K_{ijk}$, and reversing spin $i$ negates every coupling whose index set contains $i$.

The relevant subgroups of $\Aut(G)$ are those that fix both endpoints of the control task (equivalently, the groups of symmetries fixing the protocol space). Here the task is to invert the magnetization of a ferromagnetic $2\!\times\!2$ spin system from $h_i\!=\!-1$ to $h_i\!=\!+1$ for all $i$. The only nonzero multispin couplings at the protocol endpoints are the nearest-neighbor (NN) couplings $J_{01}\!=\!J_{13}\!=\!J_{23}\!=\!J_{02}\!=\!1$, with the spins labeled 0 to 3 left-to-right and top-to-bottom as in \hyperref[fig:fourSpinSymmetries]{Fig.~\ref*{fig:fourSpinSymmetries}(a)}. The automorphisms that fix both endpoints are the eight symmetries of the square, which form the dihedral group $D_4$. This group has six orbits on the 16 states [\hyperref[fig:fourSpinSymmetries]{Fig.~\ref*{fig:fourSpinSymmetries}(b)}], so $\dim M_{D_4}= 5$, and $M_{D_4}$ is parametrized by a uniform field $h$, a uniform NN coupling $J_{\rm NN}$, a uniform diagonal (next-nearest-neighbor) coupling $J_{\rm NNN}$, a uniform three-spin coupling $K$, and the four-spin coupling $Q$. These five parameters form a totally geodesic control set, and are the five parameters observed in Ref.~\cite{sawchuk_global_2026} to perform as well as optimized full control.

A controller with access to only the four local fields $h_i$ is restricted to a four-dimensional submanifold $M_{h_i} \subset \Delta^{15}$ (with $J_{\rm NN} =1$ fixed, and all other couplings set to zero). Its stabilizer subgroup~\eqref{eq:stabilizerGroup} is $\Aut(G,M_{h_i}) = D_4 \times C_2$. The $D_4$ factor permutes the spins by the symmetries of the square, and the $C_2$ factor reverses all four spins (reversing a single spin is excluded, as this would change the sign of the two associated NN couplings). The reversal interchanges the endpoints, so the largest subgroup fixing both endpoints is again $D_4$. A distribution in $M_{h_i}$ is $D_4$-invariant when the fields are, so 
\begin{equation}
    M_{D_4}\cap M_{h_i} = M_h
\end{equation}
is the one-dimensional submanifold of distributions reachable by variation of the global field $h$ alone. By \hyperref[eq:stabilizer]{Eq.~(\ref*{eq:stabilizer})}, $M_h$ is totally geodesic in $M_{h_i}$, so the optimal protocol $\eq^\star_{M_h}$ is stationary in $\Wex$ against every local-field perturbation. 

As the largest group stabilizing the endpoints, $D_4$ gives the smallest symmetric control set consistent with the task, and each of its subgroups gives a larger one. Keeping only the two diagonal reflections (generating the subgroup $C_2 \times C_2$) forces the fields to agree along each diagonal, and keeping only the left--right reflection (generating $C_2$) forces them to agree on each row:
\begin{subequations} \label{eq:fourSpinSets}
    \begin{align}
        M_{\times} &= \curly{\bm{h} \,: \, h_0 = h_3, \ h_1 = h_2}, \\
        M_{\parallel} &= \curly{\bm{h}\,:\, h_0 = h_1 , \ h_2 = h_3}.
    \end{align}
\end{subequations}
Each is the fixed-point set of a subgroup of $\Aut(G,M_{h_i})$, and hence totally geodesic in $M_{h_i}$. ($M_{\parallel}$ has a twin $\curly{\bm{h}\,:\,h_0 =h_2, \ h_1 = h_3}$ with identical geometry.) This gives a nested family with each submanifold totally geodesic in its ambient space(s): $M_{h} \subset \, M_{\parallel},M_{\times} \, \subset M_{h_i}$, i.e., $M_h$ is a submanifold of $M_{\parallel}$ and of $M_{\times}$, with both in turn submanifolds of $M_{h_i}$. The field partitions defining these control sets are shown in \hyperref[fig:fourSpinPanel]{Fig.~\ref*{fig:fourSpinPanel}}, and we return to these in \hyperref[ssec:fourspin]{Sec.~\ref*{ssec:fourspin}}.

\section{Thermodynamic curvature and local optima \label{sec:comparison}}
In this section, we show that the second variation of the excess work about a symmetric optimum splits into independent contributions from mutually orthogonal symmetry-breaking sectors~\eqref{eq:R2index}. Whether the symmetric minimum-work protocol is a minimum or a saddle point of $\Wex$ in the ambient space is then determined sector-by-sector by the curvature of the ambient space. The thermodynamic comparison theorem~\eqref{eq:R2thresholds} follows by bounding the curvature of the thermodynamic manifold along the protocol, enabling comparison to simpler, constant-curvature spaces. The representation theory and proofs are collected in \hyperref[app:secondVar]{App.~\ref*{app:secondVar}}. 

\subsection{Second variation and the index form}
Throughout, $H\subseteq \Aut(G,M')$, $M = M_H \cap M'$ is connected, $\dim M' = m'  > \dim M = m$, and $\eq^\star_M \in \mathcal{P}_M$ is a minimum-work protocol, hence a geodesic of both $M$ and $M'$ by the results of \hyperref[sec:sym]{Sec.~\ref*{sec:sym}}. Then for any one-parameter family of protocols $\eq(t;\epsilon) \in \mathcal{P}_{M'}$ with $\eq(t;0) = \eq^\star_M(t)$, the first variation of the excess work on $\mathcal{P}_{M'}$ vanishes: 
\begin{equation} \label{eq:secondVariation}
    \Wex[\eq(\cdot;\epsilon)] = \Wex^\star_M + \epsilon^2 \,I(X,X) + \mathcal{O}(\epsilon^3) \ ,
\end{equation}
for variation field $X=\partial_{\epsilon}\eq(\cdot;\epsilon)\vert_{\epsilon=0} \in T_{\eq^\star_M} \mathcal{P}_{M'}$ and
\begin{equation} \label{eq:indexForm}
    I(X,X) \equiv \int_0^1 \md t \sbrac{\, \norm{D'_tX}^2 - R'\!\paren{X, \dot{\eq}^\star_M, \dot{\eq}^\star_M, X} \,}
\end{equation}
the \textit{index form} along $\eq^\star_M$. Here $D'_t$ is the covariant derivative of $M'$ in the direction of $\dot{\eq}^\star_M(t)$, and $R'$ is the Riemann curvature tensor of $M'$, with $R'(X,\dot{\eq}^\star_M, \dot{\eq}^\star_M,X)$ at time $t$ proportional to the sectional curvature of $M'$ for the tangent plane spanned by $X(t)$ and $\dot{\eq}^\star_M(t)$ (\hyperref[app:geom]{App.~\ref*{app:geom}2,3}). 

The index form~\eqref{eq:indexForm} is half the Hessian of $\Wex$ on $\mathcal{P}_{M'}$. If $I$ is positive definite, then $\eq^\star_M$ is a local minimum, else $\eq^\star_M$ is a saddle point of $\Wex$ on $\mathcal{P}_{M'}$. (One can always deform a path to make it longer, so it can never be a local maximum.) 

\subsection{Curvature, conjugate points, and minimality: An illustration \label{ssec:RaoIllust}}
\begin{figure}[b]
    \centering
    \includegraphics[width=0.95\linewidth]{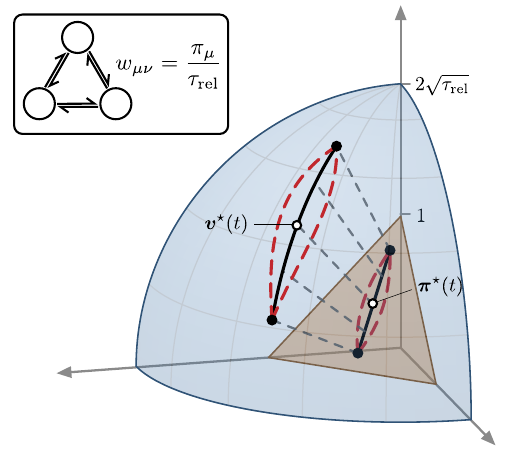}
    \caption{For transition rates $w_{\mu\nu} = \tau_{\rm rel}^{-1}\pi_\mu$ on a complete graph, the friction tensor coincides with the Fisher--Rao metric, and the embedding $v_{\mu} = 2 \sqrt{\tau_{\rm rel}\pi_{\mu}}$ maps the thermodynamic manifold $(\simp,L^+)$ onto the positive orthant of the hypersphere $S^{n}$ with all lengths preserved, shown here for three states. A geodesic on the orthant (solid black curve) is unique and minimizing, so any perturbation (dashed red curves) can only increase the excess work.}
    \label{fig:hellinger}
\end{figure}
The connection between curvature and the second variation is a classical result in differential geometry, but may be unfamiliar to a broader audience. We will therefore illustrate a system with a simple thermodynamic manifold before presenting the results in their full generality. 

Consider a system that evolves as
\begin{equation} \label{eq:MEsts}
    \dot{\bm{p}}(t) = -\tau_{\rm rel}^{-1}[\bm{p}(t)-\bm{\pi}(t)]    
\end{equation}
for some constant $\tau_{\rm rel}>0$. This is the ensemble description of Markov dynamics on a complete graph with a single degenerate relaxation timescale $\tau_{\rm rel}$ for all modes, i.e., $w_{\mu\nu} = \tau_{\rm rel}^{-1}\pi_\mu$ for all $\nu \neq \mu$. The transition rates break the locality assumption of \hyperref[ssec:setup]{Sec.~\ref*{ssec:setup}}, but they remain gauge invariant (insensitive to absolute energy scales) which is all the results below require. For these dynamics, the friction metric is the Fisher--Rao metric scaled by the lone relaxation time:
\begin{equation} \label{eq:FR}
    \dot{\eq}\t L^+ \dot{\eq} = \tau_{\rm rel}\sum_{\mu}\frac{\dot{\pi}_\mu^2}{\pi_\mu} \ .
\end{equation}
In the coordinates $v_\mu = 2 \sqrt{\tau_{\rm rel}\,\pi_\mu}$ the metric is Euclidean and, since $\eq$ is normalized, $\bm{v}$ lies on the positive orthant ($v_{\mu} \geq 0$) of a hypersphere of radius $2 \sqrt{\tau_{\rm rel}}$~\cite{ito_stochastic_2018, pistoneInformationGeometryProbability2019},
\begin{equation}\label{eq:hellinger}
    \|\dot{\eq}\|^2 =\dot{\eq}\t L^+ \dot{\eq} = | \dot{\bm{v}}|^2\, , \qquad \sum_{\mu}v_{\mu}^2 = 4\tau_{\rm rel} \ ,
\end{equation}
for standard Euclidean norm $|\cdot |$. Then if $\bm{v}^\star(t)$ traces an arc of a great circle at constant speed, $\pi^\star_{\mu}(t) = v_{\mu}^\star(t)^2 /4\tau_{\rm rel}$ traces a geodesic on the thermodynamic manifold, and hence is stationary in the excess work. \hyperref[fig:hellinger]{Figure~\ref*{fig:hellinger}} illustrates this for a three-state system. 

On a sphere of radius $r$, a geodesic arc longer than $\uppi r$ (with $\uppi$ denoting the \textit{number} pi) has continued past the antipode of its initial point, and is therefore not minimal---it has taken the long way around---and any geodesic shorter than this is the unique shortest path between its endpoints~\cite{berger_panoramic_2003}. Then a geodesic on this thermodynamic manifold is non-minimizing when $\Wex^\star = (\ell^\star)^2 > \uppi^2 / K$, where $K \equiv (4\tau_{\rm rel})^{-1}$ is the constant sectional curvature of the sphere. A geodesic on the positive orthant of the sphere can cover at most half this distance, so every geodesic of this thermodynamic manifold is a minimum-work protocol. 

The antipode of a sphere is the simplest instance of a \textit{conjugate point}. Geodesics leaving a point $\eq_0$ in slightly different directions initially separate, but positive curvature draws them back together. A point $\eq^\star(t_0)$ at which a family of neighboring geodesics rejoins $\eq^\star$ is said to be conjugate to $\eq_0$ along $\eq^\star$, just as meridians leaving the north pole of a sphere meet again at the south pole. A conjugate point at some $t \in (0,1)$ makes the index form $I$ indefinite, so $\eq^\star$ is not a local minimum, whereas a geodesic with no conjugate points in $(0,1]$ \textit{is} a local minimum~\cite{cheeger_comparison_1975}. 

Negative curvature spreads geodesics apart, so if the sectional curvature along a geodesic is nowhere positive, it has no conjugate points and is always a local minimum of the excess work. For this reason, only positive curvature is depicted in \hyperref[fig:master]{Fig.~\ref*{fig:master}(d)} and \hyperref[fig:hellinger]{Fig.~\ref*{fig:hellinger}}, as it is the only kind of curvature for which a small symmetry-breaking perturbation can possibly lower the excess work. Generally, however, thermodynamic manifolds have more complex geometry, with the sign and magnitude of the sectional curvatures varying not only from one point to the next but also between different tangent planes at a single point, so the sign of $I$ is harder to determine. We now show that $I$ can be decomposed via the dynamical symmetries, making the problem more tractable and enabling the identification of symmetries (if any) that are advantageous to break.

\subsection{Decomposition of the index form by symmetry}
Since $M$ is totally geodesic in $M'$, the covariant derivative $D'_t$ of $M'$ and the curvature operator $X \mapsto R'(X,\dot{\eq}^\star_M)\dot{\eq}^\star_M$ map tangent vectors of $M$ to tangent vectors of $M$, and by orthogonality, normal vectors to normal vectors (\hyperref[app:comparison]{App.~\ref*{app:comparison}}). The index form therefore decouples,
\begin{equation} \label{eq:Isplit}
    I(X,X) = I(X^{\parallel}, X^{\parallel}) + I(X^{\perp}, X^{\perp})\ .
\end{equation}
Here $X \in T_{\eq^\star_M}\mathcal{P}_{M'}$ is a perturbation of an $M$-geodesic in $M'$, and $X^{\parallel}$ and $X^{\perp}$ are its respective components tangent and normal to $M$. Since $\eq^\star_M$ is a minimum-work protocol on $\mathcal{P}_M$ by assumption, the tangential part $I(X^{\parallel},X^{\parallel})$ is nonnegative. A symmetric optimum can therefore fail to be a minimum in $M'$ only through the $m_{\perp} \equiv m'-m$ normal (symmetry-breaking) directions, one per lifted constraint, and it suffices to test only those.

The index form may be decomposed further still. On the symmetric submanifold, the equivariance of the Laplacian~\eqref{eq:isocriterion} reduces to commutation,
\begin{equation} \label{eq:commute}
    \sigma \in H , \ \eq\in M_H \implies [L(\eq),P_\sigma] = 0  \  ,
\end{equation}
and the same holds for the metric $L^+$. The covariant derivative and curvature operator are determined by the metric alone, so they inherit its equivariance and likewise commute with every isometry $P_{\sigma}$ along $\eq^\star_M$~(\hyperref[app:comparison]{App.~\ref*{app:comparison}}).

Just as a Hamiltonian that commutes with a group of operators is block diagonal in a basis adapted to that group, the space of perturbations splits into mutually $L^+$-orthogonal \textit{sectors} $U_\rho$, one for each real irreducible representation (irrep) $\rho$ of $H$. The sectors cannot be mixed by any operator that commutes with every $P_{\sigma}$, so the second variation decouples sector by sector. The orthogonal projectors $\Pi_{\rho}$ onto the sectors are obtained from $H$ alone (\hyperref[app:sectors]{App.~\ref*{app:sectors}}), so they can be computed once for the entire protocol. By contrast, the relaxation modes of the dynamics diagonalize the metric at each point~\cite{sawchuk_global_2026} but change along the protocol and generally do not diagonalize the curvature operator. The natural coordinates for symmetry breaking are therefore derived from the irreps, not the relaxation modes.

The irrep $\rho$ describes how symmetries in $H$ act on perturbations in $U_{\rho}$: which permutations reverse the perturbations, which mix them among themselves, and which leave them unchanged. In particular, $U_0$ (with $\rho\!=\!0$ the trivial irrep) is the \textit{symmetric sector} consisting of the directions fixed by every $P_{\sigma}$, and its normalization-conserving part is the tangent space to the symmetric submanifold $T_{\eq}M_H = U_0 \cap \one^\perp$. The orbit-average operator $\Pi_0$ is its projector. The remaining sectors ($\rho\!\neq\!0$) are the \textit{symmetry-breaking sectors}, containing no directions fixed by all of $H$. 

The sectors $U_0,U_1,\dots$ decompose all of $\mathbb{R}^{n+1}$ and so contain any probability-conserving perturbation of a symmetric protocol. A controller restricted to $M'$ can only reach the part of each sector lying in $T_{\eq}M'$, the \textit{accessible sector} $\mathcal{U}_{\rho} \equiv U_{\rho}\cap T_{\eq}M'$ (in particular, $T_{\eq}M= \mathcal{U}_0$). The symmetry-breaking perturbations available to the controller $M'$ thus decompose into one orthogonal component per pattern of symmetry breaking
\begin{equation} \label{eq:Ksectors}
    N_{\eq}M = \bigoplus_{\rho \neq 0} \mathcal{U}_\rho \subset T_{\eq}M' \ .
\end{equation}
If no nontrivial directions of $U_{\rho}$ are accessible in $T_{\eq}M'$ ($\mathcal{U}_{\rho} = \curly{0}$), the controller cannot break the corresponding symmetry. We will see an example of this in \hyperref[ssec:fourspin]{Sec.~\ref*{ssec:fourspin}}.

Define a parallel orthonormal frame $\curly{\bm{n}_I^{\rho}}_{I=1}^{\dim \mathcal{U}_{\rho}}$ for each sector $\mathcal{U}_{\rho}$ along $\eq^\star_M$ [\hyperref[eq:frame]{Eq.~(\ref*{eq:frame})}]. A variation field in sector $\rho$ decomposes as $X_{\rho}(t) = x^I_{\rho}(t) \bm{n}^{\rho}_I(t)$ (summation over repeated indices implied), with $\bm{x}_{\rho}(0) = \bm{x}_{\rho}(1) = 0$. Capital indices $I,J,\dots$ run over frame vectors, as distinct from Greek indices over states and lowercase Latin indices over control parameters. Because the frame is parallel, $D'_t X_{\rho} = \dot{x}^I_{\rho} \bm{n}^{\rho}_I$, and 
\begin{equation}\label{eq:Ioscillator}
    I(X_{\rho},X_{\rho}) = \int_0^1 \md t \,\big[|\dot{\bm{x}}_{\rho}|^2 - \bm{x}_{\rho}\t(\Wex^\star_M \kcur_{\rho} ) \bm{x}_{\rho}\big] \ ,
\end{equation}
for Euclidean norm $|\cdot |$ on the frame coefficients $\bm{x}_{\rho}(t)$ and \textit{sector curvature operators} [\hyperref[fig:master]{Fig.~\ref*{fig:master}(c)}]
\begin{equation} \label{eq:Krho}
    \paren{\kcur_\rho(t)}_{IJ}
    \equiv \frac{R'\paren{\bm{n}^\rho_I, \dot{\eq}^\star_M,
    \dot{\eq}^\star_M, \bm{n}^\rho_J}}{\|\dot{\eq}^\star_M \|^2} \ , \quad
    \bm{n}^\rho_I \in \mathcal{U}_\rho \ .
\end{equation}
\hyperref[eq:Ioscillator]{Equation~(\ref*{eq:Ioscillator})} is the action of an oscillator with time-dependent stiffness matrix $\Wex^\star_M\kcur_{\rho}(t)$, and a return of this oscillator to the origin before the protocol ends marks a conjugate point. This gives an alternative perspective on the effect of positive and negative curvatures: Positive curvature acts as a restoring force that draws perturbed protocols back toward $\eq^\star_M$, and negative curvature acts as a repulsive force. Summing~\eqref{eq:Ioscillator} over sectors gives the decomposition \hyperref[eq:R2index]{Eq.~(\ref*{eq:R2index})} of the second variation.

A further simplification applies within each sector, though the details are rather technical so are largely deferred to \hyperref[app:sectors]{App.~\ref*{app:sectors}}. In brief, the smallest $H$-invariant subspaces of $\mathcal{U}_{\rho}$ all have dimension $d_{\rho}$ (the \textit{irrep dimension}) and $\mathcal{U}_{\rho}$ is the direct sum of $m_{\rho}$ of them (the \textit{multiplicity}), so $\dim \, \mathcal{U}_{\rho} = m_\rho d_{\rho}$. For real-type irreps (defined in \ref{app:sectors}), operators that commute with every $P_{\sigma}$ can mix the $m_{\rho}$ subspaces, but act identically on the $d_{\rho}$ directions, so the oscillator of sector $\mathcal{U}_\rho$ separates into $d_{\rho}$ identical and independent oscillators of dimension $m_{\rho}$. This applies to the four-spin system, for which the normal space of $M_{D_4}$ in $\Delta^{15}$ is ten-dimensional and splits into three symmetry-breaking sectors with $(d_{\rho},m_{\rho})=(1,3),(2,3),(1,1)$. This can be seen in the diagonalized matrices plotted in \hyperref[fig:master]{Fig.~\ref*{fig:master}(c)}, where the two red $3\times 3$ blocks are identical. The ten-dimensional test thus reduces to three independent tests of dimensions $m_{\rho}=3,3,1$. More generally, if not all irreps are real, the oscillator of sector $\mathcal{U}_{\rho}$ separates into $d_{\rho}/e_{\rho}$ identical oscillators of dimension $e_{\rho}m_{\rho}$ with $e_{\rho} \in \curly{1,2,4}$ depending on the irrep $\rho$ [\hyperref[eq:equivariantForm]{Eq.~(\ref*{eq:equivariantForm})}].

\subsection{Jacobi equation and the thermodynamic comparison theorem}
The Euler--Lagrange equation for the oscillator~\eqref{eq:Ioscillator} of sector $\rho$ is
\begin{equation} \label{eq:jacobiOsc}
    \ddot{\bm{x}}_\rho(t) + \Wex^\star_M\,\kcur_\rho(t)\,\bm{x}_\rho(t) = 0 \ , 
\end{equation}
the \textit{Jacobi equation}~\eqref{eq:jacobi} of $M'$ restricted to sector $\rho$ written in the parallel frame. Its solutions with $\bm{x}_{\rho}(0)=0$ are the Jacobi fields of sector $\rho$. They give the first-order separation between $\eq^\star_M$ and the neighboring geodesics of $M'$ that leave $\eq_0$ with velocity deflected into sector $\rho$. (The Jacobi fields need not vanish at $t=1$, so they are not in $T_{\eq^\star_M}\mathcal{P}_{M'}$ in general.) The second variation with respect to symmetry-breaking perturbations in symmetry sector $\rho$ is therefore positive definite if and only if no nontrivial solutions of~\eqref{eq:jacobiOsc} initiated at the origin return there before the end of the protocol. Conversely, the presence of a conjugate point in \textit{any} sector makes $I$ indefinite, and the sectors with conjugate points indicate the symmetries worth breaking~\cite{cheeger_comparison_1975, Note2}.\footnotetext{In the language of Morse theory, the number of conjugate points in $(0,1)$ summed over the sectors is equal to the \textit{Morse index} of the critical point $\eq^\star_M \in \mathcal{P}_{M'}$, the dimension of the largest subspace of $T_{\eq^\star_M}\mathcal{P}_{M'}$ on which $I$ is negative definite~\cite{cheeger_comparison_1975}.}

Locating conjugate points requires solving \hyperref[eq:jacobiOsc]{Eq.~(\ref*{eq:jacobiOsc})}, which generally cannot be done in closed form. The comparison theorem avoids this by bounding $\kcur_{\rho}(t)$ with its extremal eigenvalues along the protocol,
\begin{subequations}\label{eq:kminmax}
    \begin{align}
        k^{\max}_\rho &\equiv \sup_{t\in[0,1]} \lambda_{\max}\sbrac{\kcur_\rho(t)} \ , \\
        k^{\min}_\rho &\equiv \inf_{t\in[0,1]} \lambda_{\min}\sbrac{\kcur_\rho(t)} \ .
    \end{align}
\end{subequations}
By the Sturm comparison principle (a stiffer spring returns to the origin no later than a softer one~\cite{berger_panoramic_2003}), the first return of a Jacobi field in sector $\rho$ occurs no sooner than the first zero of $\ddot{x} + \Wex^\star_M k^{\max}_{\rho}x=0$, and no later than the first zero of $\ddot{x} + \Wex^\star_M k^{\min}_{\rho}x=0$. An oscillator with constant stiffness initiated at the origin returns to it before $t=1$ only if its stiffness is greater than $\uppi^2$, thus $\eq^\star_M$ cannot have a conjugate point in sector $\rho$ if $\Wex^\star_M k^{\max}_{\rho} < \uppi^2$; conversely, it must if $\Wex^\star_M k^{\min}_{\rho} > \uppi^2$. Since $\eq^\star_M$ is a local minimum of $\Wex$ on $\mathcal{P}_{M'}$ if no sector has a conjugate point, and is a saddle point otherwise, these bounds give the \textit{thermodynamic comparison theorem}: 
\begin{subequations} \label{eq:R2thresholds} 
    \begin{align}
        &k^{\max}_\rho < \frac{\uppi^2}{\Wex^\star_M} \ \ \forall \,\rho  &&\implies \text{local minimum} \label{eq:R2min} \\
        &k^{\min}_\rho > \frac{\uppi^2}{\Wex^\star_M} \ \ \text{for some }\rho &&\implies \text{saddle point} \label{eq:R2max} \ .
    \end{align}
\end{subequations}
Geometrically, the theorem compares the thermodynamic length $\ell^\star = \sqrt{\Wex^\star_M}$ of $\eq^\star_M$ with the pole-to-pole distance $\uppi/\sqrt{k}$ on two comparison spheres of curvatures $k = k_{\rho}^{\min}$ and $k_{\rho}^{\max}$. This is a special case of the Rauch comparison theorem~\cite{berger_panoramic_2003, cheeger_comparison_1975}. A sector with negative curvature everywhere along the protocol satisfies \hyperref[eq:R2min]{Eq.~(\ref*{eq:R2min})}, so no local perturbation in that sector can reduce the excess work, while a sector with sufficiently large positive curvature indicates a symmetry worth breaking. Between these thresholds, the test is inconclusive and conjugate points must be located by direct integration of \hyperref[eq:jacobiOsc]{Eq.~(\ref*{eq:jacobiOsc})} (\hyperref[app:numerics]{App.~\ref*{app:numerics}}). In the case of a single degenerate timescale on a complete graph (as in \hyperref[fig:hellinger]{Fig.~\ref*{fig:hellinger}}), $k^{\min} = k^{\max} = K$, and the test is sharp.

\begin{figure*}
    \includegraphics[width=\linewidth]{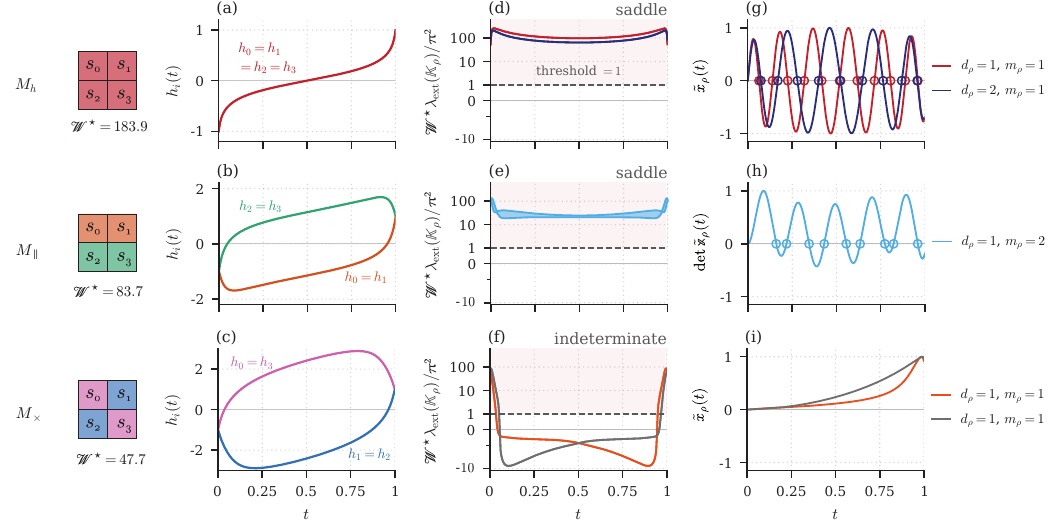}
    \caption{
    Field control of the four-spin system on the nested control sets $M_h \subset \, M_{\parallel}, M_{\times} \, \subset M_{h_i}$, one per row. Each control set is defined by the field partition shown at the left, with spins of the same color sharing a field. 
    (a, b, c) Minimum-work protocols in the coordinates $h_i(t)$, colored by field group. Swapping the colors in (b) or (c) gives a second minimum-work protocol in the same control set.
    (d, e, f) Curvature along each protocol in the ambient $M_{h_i}$, shown for each symmetry-breaking sector of the protocol's symmetry group. Each curve is a scaled extremal curvature $\Wex^\star \lambda_{\rm ext}[\kcur_{\rho}(t)]/\uppi^2$, so that the thresholds of \hyperref[eq:R2thresholds]{Eq.~(\ref*{eq:R2thresholds})} sit at unity (dashed line). In (d) and (f), every sector contains just one copy of its irrep ($m_{\rho}=1$) so $\kcur_{\rho}$ has a single distinct eigenvalue. The sector in (e) has $m_{\rho}=2$, and the blue shaded band is bounded by its two eigenvalues at each $t$. The scaled curvature of any plane spanned by the protocol velocity and a direction in that sector lies within this band.
    (g, h, i) Conjugate points (circles) from direct integration of the sector Jacobi equation~\eqref{eq:jacobiOsc}. This reduces to a scalar equation in (g) and (i), where the normalized solution $\tilde{x}_{\rho}(t) = x_{\rho}(t) / \max_t|x_{\rho}(t)|$ is shown. For the $m_{\rho}=2$ sector of (h), conjugate points are the zeros of the normalized determinant of the matrix $\mathbbm{x}_{\rho}(t)$ of Jacobi fields [\hyperref[eq:jacMatrix]{Eq.~(\ref*{eq:jacMatrix})}].
    }
    \label{fig:fourSpinPanel}
\end{figure*}

\subsection{Example: Field control of four spins\label{ssec:fourspin}}
We now apply the comparison theorem to the four-spin ferromagnetic system of \hyperref[ssec:fourSpinSym]{Sec.~\ref*{ssec:fourSpinSym}} controlled only by local fields ($M' = M_{h_i}$). All group-theoretic and geometric quantities and plots are reproducible with the accompanying code~\cite{code_ref}, with additional details in \hyperref[app:numerics]{App.~\ref*{app:numerics}}.

Recall that the the task is to invert the magnetization $\bm{h}=-\one\to +\one$ at fixed nearest-neighbor couplings $J_{\rm NN} = 1$, and the largest subgroup of $\Aut(G,M_{h_i})$ that fixes the endpoints is $D_4$. Then $M_{D_4} \cap M_{h_i} = M_{h}$ is the one-dimensional submanifold for control of a global field $h$. Any symmetry breaking within $M_{h_i}$ corresponds to increased spatial resolution of the field control, and the sector decomposition can be given in field coordinates. The space of field perturbations $\delta \bm{h}$ splits as $\mathbb{R}^4 = \mathcal{U}_0 \oplus \mathcal{U}_1 \oplus \mathcal{U}_2$, with
\begin{subequations}
    \begin{align}
        \mathcal{U}_0 &= \Span\curly{(1,1,1,1)\t} \, , \\
        \mathcal{U}_1 &= \Span\curly{(1,-1,-1,1)\t} \, ,  \\
        \mathcal{U}_2 &= \Span\curly{(1,1,-1,-1)\t, \, (1,-1,1,-1)\t}  \, . 
    \end{align}
\end{subequations}
Here, the dimensions of the irreps $\rho = 0,1,2$ are $d_{\rho}=1,1,2$, and their multiplicities in the accessible sectors of $M_{h_i}$ are $m_{\rho} = 1,1,1$. $\mathcal{U}_0 = T_{\bm{h}}M_h$ is the global field, and the symmetry-breaking sectors are split into field perturbations with opposite signs on the two diagonals ($\mathcal{U}_1$) and field perturbations with opposite signs on the two rows or the two columns ($\mathcal{U}_2$). These are the field-controllable directions within the sectors $U_1$ and $U_2$ of $M_{D_4}$ in $\Delta^{15}$ [\hyperref[fig:master]{Fig.~\ref*{fig:master}(b),(c)}]. The third symmetry-breaking sector, $U_3$, comprises perturbations that raise or lower the energies of the zero net magnetization states with aligned rows relative to those with aligned columns [sets outlined in blue in \hyperref[fig:master]{Fig.~\ref*{fig:master}(b)}], leaving every other symmetry intact. No field perturbation realizes this pattern, so $\mathcal{U}_3 = \curly{\bm{0}}$. In the coupling coordinates, $U_3$ alternates the sign of the couplings on the horizontal and vertical bonds, and so requires control of the couplings $J_{ij}$. 

In terms of these sectors, the control sets of \hyperref[eq:fourSpinSets]{Eq.~(\ref*{eq:fourSpinSets})} have tangent spaces
\begin{subequations}
\label{eq:fourSpinTangentSpaces}
    \begin{align}
        T_{\bm{h}}M_{\times} &= \mathcal{U}_0 \oplus \mathcal{U}_1 \, , \\
        T_{\bm{h}}M_{\parallel} &= \mathcal{U}_0 \oplus \Span\curly{(1,1,-1,-1)\t}.
    \end{align}
\end{subequations}
$M_{\times}$ breaks the symmetry fully in sector $\mathcal{U}_1$, while $M_{\parallel}$ fixes one direction of $\mathcal{U}_2$.

The results are summarized in \hyperref[fig:fourSpinPanel]{Fig.~\ref*{fig:fourSpinPanel}}. Direct optimization (procedure and data from Ref.~\cite{sawchuk_global_2026}) gives $\Wex^\star_{M_h}=183.9$, $\Wex^\star_{M_{\parallel}} = 83.7$, and $\Wex^\star_{M_{\times}}=\Wex^\star_{M_{h_i}}=47.7$ in arbitrary units. The protocols found on $M_{\times}$ and $M_{h_i}$ are identical to numerical precision. The comparison theorem anticipates part of this hierarchy from the one-parameter protocol $h^\star$ alone. Along $h^\star$, the scaled curvature $\Wex^\star k^{\min}_\rho / \uppi^2$ exceeds unity in both accessible sectors [\hyperref[fig:fourSpinPanel]{Fig.~\ref*{fig:fourSpinPanel}(d)}], so the saddle-point condition~\eqref{eq:R2max} guarantees that refining the field control in either pattern lowers the cost. The theorem does not say definitively which refinement lowers it more, but the greater curvature in $\mathcal{U}_1$ and the corresponding earlier arrival of the first conjugate point [\hyperref[fig:fourSpinPanel]{Fig.~\ref*{fig:fourSpinPanel}(g)}] are consistent with the observed $\Wex^\star_{M_{\times}}< \Wex^\star_{M_{\parallel}}$.

The manifold $M_{\parallel}$ [\hyperref[fig:fourSpinPanel]{Fig.~\ref*{fig:fourSpinPanel}(b,e,h)}] has the symmetry group $C_2$ generated by the left--right reflection, under which $\mathcal{U}_1$ and the normal direction of $\mathcal{U}_2$ merge into a single two-dimensional sector ($d_{\rho}=1, m_{\rho}=2$). Its curvature exceeds the threshold, so this protocol is also a saddle point of $\Wex$ on $\mathcal{P}_{M_{h_i}}$, consistent with $\Wex_{M_{\parallel}}^\star > \Wex_{M_{h_i}}^\star$. 

The optimal protocol of $M_{\times}$ [\hyperref[fig:fourSpinPanel]{Fig.~\ref*{fig:fourSpinPanel}(c,f,i)}] has symmetry group $C_2\times C_2 \subseteq D_4$, generated by the two diagonal reflections of the square, under which $\mathcal{U}_2$ splits into one-dimensional sectors spanned by the perturbation patterns $(1,0,0,-1)\t$ and $(0,1,-1,0)\t$. Their curvatures are negative over most of the protocol, but positive spikes near the endpoints leave the test inconclusive [\hyperref[fig:fourSpinPanel]{Fig.~\ref*{fig:fourSpinPanel}(f)}]. Direct integration of the Jacobi equation shows that no Jacobi field returns to the origin in $(0,1]$ [\hyperref[fig:fourSpinPanel]{Fig.~\ref*{fig:fourSpinPanel}(i)}], so the optimal protocol on $\mathcal{P}_{M_{\times}}$ is indeed a local minimum of $\Wex$ on $\mathcal{P}_{M_{h_i}}$.

\section{Convexity and global optimality \label{sec:convexity}}
Here we show that an optimal protocol on a symmetric submanifold is globally optimal in the ambient space when the protocol space is convex and the conductance on every edge is a concave function of the equilibrium probabilities of the states it joins. As before, let $M' \subseteq \simp$ be a control set, $H \subseteq \Aut(G,M')$, and $M = M_H \cap M'$. The projector onto the symmetric sector $U_0$ of $H$ is the orbit-average operator introduced in \hyperref[eq:R3]{Eq.~(\ref*{eq:R3})}, and it can be written as an average over the group,
\begin{equation} \label{eq:Pi0}
    \Pi_0 = \frac{1}{\abs{H}}\sum_{\sigma \in H} P_\sigma \ .
\end{equation}
Acting on a distribution, $\Pi_0$ sends the probability mass of each state to the mean probability of its $H$-orbit, and so erases every symmetry-breaking component. If the protocol space $\mathcal{P}_{M'}$ is convex, then for any protocol $\eq \in \mathcal{P}_{M'}$, $\Pi_0 \eq \in \mathcal{P}_{M}$ is the symmetrized protocol for the same task. So if $\Wex$ is also a convex functional, then
\begin{subequations} \label{eq:jensen}
    \begin{align}
        \Wex\sbrac{\Pi_0 \eq} &\leq \frac{1}{\abs{H}}\sum_{\sigma \in H}\Wex\sbrac{P_{\sigma}\eq} \\
        &= \Wex[\eq] \ .
    \end{align}
\end{subequations}
The first step is Jensen's inequality and the second follows from isometry ($\Wex[P_{\sigma}\eq] = \Wex[\eq]$). Hence for convex excess work and convex protocol space $\mathcal{P}_{M'}$, a minimum-work protocol on $\mathcal{P}_M$ is a minimum-work protocol on $\mathcal{P}_{M'}$, and the cost of constraint~\eqref{eq:costOfConstraint} vanishes: $\Delta\Wex(M\!\to\!M')=0$. 

Convexity of the excess work is determined by the kinetics, while convexity of the protocol space is determined by the control set. To see where the kinetics enter, write the integrand of $\Wex$ as a Dirichlet form and take its Legendre dual~\cite{boydConvexOptimizationStephen2004}:
\begin{subequations} \label{eq:39full}
    \begin{align}
        \dot{\eq}\t L^+ \dot{\eq} &= \sum_{(\mu,\nu) \in \mathcal{E}} \cond_{\mu\nu} \paren{\phi_\mu - \phi_\nu}^2  \label{eq:dirichlet} \\&=  \sup_{\bm{\psi}} \Big\{\, 2\bm{\psi}\t\dot{\eq} - \sum_{e\in\mathcal{E}} \cond_e(\eq)\paren{\psi_\mu - \psi_\nu}^2 \Big\} \ . \label{eq:legendreMain}
    \end{align}
\end{subequations}

The potential $\bm{\phi}$, defined as the solution to $\dot{\eq}=L\bm{\phi}$, is the negative relative lag under linear response, $\phi_\mu \simeq - (p_\mu - \pi_\mu)/\pi_\mu$~\cite{sawchuk_thermodynamic_2026}, and \hyperref[eq:39full]{Eq.~(\ref*{eq:39full})} is the power dissipated in an electrical network with conductance $\cond_{\mu\nu}$ on edge $(\mu,\nu)$. The supremum in \hyperref[eq:legendreMain]{Eq.~(\ref*{eq:legendreMain})} is attained at $\bm{\psi}=\bm{\phi} = L^+ \dot{\eq}$. Integrating over $t$ and taking the supremum preserves convexity, so $\Wex$ is a convex functional when the argument of the supremum in~\eqref{eq:legendreMain} is jointly convex in $(\eq,\dot{\eq})$. For fixed $\bm{\psi}$, the first term is linear (hence convex) in $\dot{\eq}$, and the second term is convex in $\eq$ if each $\cond_{\mu\nu}$ is jointly concave in $(\pi_\mu,\pi_\nu)$. Thus $\Wex$ is a convex functional when the conductance is concave. 

The protocol space $\mathcal{P}_{M'}$ is convex if $\alpha \eq + (1-\alpha)\eq' \in \mathcal{P}_{M'}$ for all $\eq, \eq' \in \mathcal{P}_{M'}$ and $\alpha \in [0,1]$. Since $\simp$ is convex (a mixture of two distributions is a distribution), so is $\mathcal{P}_{\simp}$. Thus for concave conductances, \hyperref[eq:jensen]{Eq.~(\ref*{eq:jensen})} holds for full control, and optimal control of the lumped chain is optimal control of the system, $\Wex^\star_{M_H} = \Wex^\star_{\simp}$. A constrained control set $M'$ is generally not convex, so $\eq \in \mathcal{P}_{M'}$ does not imply $\Pi_0 \eq \in \mathcal{P}_{M'}$. Under concave kinetics, then, symmetry breaking can offer an advantage only when one cannot fully control the lumped chain (i.e., $M_H \not\subset M'$) and its protocol space is nonconvex.

What is the physical content of the concavity of the conductance $\cond_e$, the one-way equilibrium flux across an edge? Under the assumptions of \hyperref[ssec:setup]{Sec.~\ref*{ssec:setup}}, each rate is a function of $\e{-(V_\mu - V_\nu)} = \pi_\mu/\pi_\nu$, the Boltzmann factor of the energy difference  across its edge, so $\cond_{\mu\nu} = \pi_\nu w_{\mu\nu}(\pi_\mu/\pi_\nu)$ is jointly concave in $(\pi_\mu,\pi_\nu)$ if and only if $w_{\mu\nu}$ is concave in $\pi_\mu/\pi_\nu$~\cite{boydConvexOptimizationStephen2004}. Concavity of the conductance is thus a condition of diminishing returns in the driving force. The rate must grow no faster than linearly in $\e{-(V_\mu - V_\nu)}$, so that the increased transition rate from destabilizing a state is outweighed by the decreased equilibrium probability in that state, and the flux cannot be increased indefinitely by tilting the energy landscape. Common rate laws satisfy this condition: Glauber rates $w_{\mu\nu} = [1+(\pi_\mu/\pi_\nu)^{-1}]^{-1}$ saturate, and while square-root rates $w_{\mu\nu} = \sqrt{\pi_\mu/\pi_\nu}$ grow without bound, the flux $\sqrt{\pi_\mu\pi_\nu}$ is largest when $\pi_\mu=\pi_\nu$ (for fixed $\pi_\mu + \pi_\nu$).  

The same structure appears in discrete optimal transport. There, the edge weights are typically generalized means of the adjacent probability masses, and their concavity is the standard assumption under which the transport action is convex and geodesics exist~\cite{maas_gradient_2011, erbarRicciCurvatureFinite2011, erbarGeometryGeodesicsDiscrete2019}. This parallel is expected, since the thermodynamics of control and optimal transport are closely related~\cite{van_vu_thermodynamic_2023, yoshimuraHousekeepingExcessEntropy2023,zhongLinearResponseEquivalence2024, sawchuk_thermodynamic_2026}. 

\section{Discussion \label{disc}}
The cost of constraint $\Delta \Wex(M\!\to\!M')$ is a difference between two optima, and computing it generally means solving both optimization problems. Under certain conditions on the kinetics and control space, we have shown that it vanishes for parameters adapted to the symmetries of the task, which can be identified from the transition network prior to any optimization. The more general question is which refinements of control are most beneficial, or equivalently, which constraints are most costly. Whether a given refinement can reduce the excess work is answered locally by the extrinsic geometry, quantified by the second fundamental form $\II$ of $M$ in $M'$~\eqref{eq:IIgrad} along a constrained optimum $\eq^\star_M$. If $\II \neq 0$, refining the control lowers the cost. If $\II =0$, as for symmetric constraints, the question passes to the second variation, which is determined by the curvature of the ambient space along the protocol. Its sign is decided sector by sector, either by locating conjugate points directly or, more cheaply, by the thermodynamic comparison theorem, which needs only the extremal curvatures along the protocol.

Still, it remains to determine \textit{how} costly a constraint is. Van Vu and Saito give a lower bound on the excess work for a given control set prior to any optimization~\cite{van_vu_thermodynamic_2023}, but a lower bound does not by itself resolve $\Delta \Wex (M\!\to\!M')$. Such bounds would allow the parameters of a control set to be screened in advance, and ranked whenever their bounds do not overlap. Here, we've used curvature only to determine whether a protocol is a local minimum, but comparison geometry offers further machinery that may be of use. This motivates a closer study of the curvature of thermodynamic manifolds and its dependence on the rate law and on features of the transition network, such as its cycles, connectivity, and automorphisms. Such a study could also sharpen the thermodynamic comparison theorem. Replacing the spaces of constant curvature that set its thresholds with the thermodynamic manifolds of different, perhaps simpler, transition networks, whose curvatures along the protocol bound those of the manifold of interest, could tighten the thresholds and lead to analytic insights. Further, the comparison theorem concerns only the case $\II =0$. It is natural to expect that the magnitude $\|\II\|$ of the extrinsic curvature relates to the cost of constraint, though this magnitude alone cannot determine it. There may also be more to say beyond linear response: We show in \hyperref[ssec:exactExcess]{App.~\ref*{ssec:exactExcess}} that the sector decomposition does not depend on slow driving, since the second variation of the exact excess work at a symmetric protocol is $H$-equivariant and hence block-diagonal across the sectors. The curvature interpretation and comparison theorem are lost, however. 

The framework presented here is not completely general, and we close by considering how its assumptions might be relaxed.

\paragraph{Bounded controls.} 
The thermodynamic manifold is typically incomplete, in that for some common rate laws a geodesic can reach the boundary of the simplex after a finite thermodynamic length (\hyperref[app:global]{App.~\ref*{app:global}}). For some endpoints, then, the infimum of the excess work may be attained only by protocols that empty a state completely, requiring nonphysical infinite state energies. Optimal protocols demanding such extreme parameters have been reported in thermodynamic geometry near second-order phase transitions, where the minimum-dissipation path can pass through infinite temperature~\cite{basri_thermodynamic_2025}. No laboratory controller is so unconstrained, however, and bounding the state energies $\bm{V}$ to a closed box restricts the accessible distributions to a compact region of $\Delta^n$ on which minimum-work protocols exist between any two points. The caveat is that an optimal protocol may saturate the bounds and run along the boundary of the region, in which case it is not a geodesic and the local tests of \hyperref[sec:comparison]{Sec.~\ref*{sec:comparison}} do not apply. Still, the convexity argument of \hyperref[sec:convexity]{Sec.~\ref*{sec:convexity}} requires only that the constrained region of $\simp$ be convex and conform to the symmetries of the task. A uniform bound on the energy differences $|V_{\mu}-V_{\nu}|$, for example, defines such a region, on which the global optimum is attained and symmetric. 

\paragraph{Rates with a fixed reference energy.} The assumption that rates depend only on energy differences excludes rate laws sensitive to absolute energies, such as Arrhenius rates $w_{\mu\nu} = \e{-(E^\ddagger_{\mu\nu}-V_\nu)}$ over fixed barriers $E^{\ddagger}_{\mu\nu}=E^{\ddagger}_{\nu\mu}$. A uniform shift of the state energies leaves $\eq$ unchanged and costs no work beyond the total free-energy change fixed by the task. But raising every well against fixed barriers is equivalent to lowering every barrier, rescaling the transition rates: $\rma(\bm{V} + c\one) = \e{c}\,\rma(\bm{V})$. The conductances are $\cond_{\mu\nu} = \e{-E^{\ddagger}_{\mu\nu}+F}$ with $F$ the equilibrium free energy, so
\begin{equation} \label{eq:arrheniusMetric}
    L^+ = \e{-F} L_0^+  \ ,
\end{equation}
for constant Laplacian $L_0$ of barrier weights $\e{-E^\ddagger_{\mu\nu}}$. If the shift $\bm{V} \mapsto \bm{V}+ c \one$ is available to the controller, the friction can be made arbitrarily small at no additional cost, and the minimum-work problem is ill-posed. This occurs for any Markov jump process with unconstrained transition rates; a known resolution is to fix the time-averaged mobility~\cite{van_vu_thermodynamic_2023}. A natural remedy here is to bound the state energies, since Arrhenius rates presume barriers high above the wells ($E^\ddagger_{\mu\nu} - V_\nu \gg 1$). The optimal protocol should then be expected to saturate the bound, holding the wells as high as permitted. Alternatively, the formalism of this paper can be recovered with the conformally flat metric~\eqref{eq:arrheniusMetric} by specifying a `gauge' $F(\bm{V}) = F(\eq)$ so that $\bm{V}\leftrightarrow \eq$ is a bijection. Calling it a gauge is an abuse of terminology, as such a choice has physical consequences. An automorphism $\sigma$ of the Markov graph (with edges colored by barrier height as in \hyperref[app:colored]{App.~\ref*{app:colored}}), for instance, is an isometry only if it also preserves the gauge, $F(P_\sigma \eq) = F(\eq)$. The gauge $\sum_\mu V_\mu = 0$ gives $\e{F} = \big(\prod_{\mu}\pi_\mu\big)^{\frac{1}{n+1}}$, which is permutation invariant, so every automorphism remains an isometry. Pinning a reference state $V_0 = 0$ instead gives $\e{F}=\pi_0$, and only automorphisms that fix state $0$ remain isometries. In both gauges, $\e{F}$ is concave in $\eq$, so the convexity result of \hyperref[sec:convexity]{Sec.~\ref*{sec:convexity}} carries over, though this is not guaranteed for an arbitrary gauge.

\paragraph{Beyond detailed balance.} For systems driven between nonequilibrium steady states, a tractable cost function is the excess (nonadiabatic) entropy production of the Hatano--Sasa decomposition~\cite{oonoSteadyStateThermodynamics1998, hatanoSteadyStateThermodynamics2001, espositoThreeFacesSecond2010}. Under slow driving, the rate of this entropy production is the quadratic form $\dot{\eq}\t\mathrm{Sym}(L^+_{\rm fwd})\dot{\eq}$ with $L_{\rm fwd} = -\rma \Deq$ a nonsymmetric matrix for instantaneous stationary distribution $\eq$~\cite{mandalAnalysisSlowTransitions2016, matsuoAnalysisEntropyProduction2022}. Provided that no accessible control direction changes the rates at fixed stationary distribution, $\Sym(L^+_{\rm fwd})$ is again a Riemannian metric on the simplex. The isometry condition is unchanged in form, now requiring equivariance of $\Sym(L^+_{\rm fwd})$ under $P_{\sigma}$, for which equivariance of $L_{\rm fwd}$ remains sufficient. Dynamical symmetries of the driven system are therefore isometries as before, so symmetric optima remain stationary points of the cost on the protocol space, and the sector decomposition and comparison theorem of \hyperref[sec:comparison]{Sec.~\ref*{sec:comparison}} apply as written. 

The converse fails, however. Since $\Sym[(L_{\rm fwd}\t)^+]=\Sym(L^+_{\rm fwd})$, the cost of a protocol depends on the currents but not on their direction, so a permutation that maps the dynamics onto its time reversal (such as a reflection of a driven ring) can be an isometry without being a dynamical symmetry. The assumption that the stationary distribution determines the rates is somewhat limiting, and the more general case is left to future work. Moreover, the excess entropy production omits the dissipation that sustains the steady state, and a strong argument has been made that the natural cost function is instead the total heat dissipated during the protocol and the subsequent relaxation to the final steady state~\cite{davisActiveMatter, sorianiControlActiveField2025}, a choice that fundamentally changes the control problem.

\begin{acknowledgments}
We thank Antonio Patr\'{o}n Castro and John Bechhoefer (Simon Fraser University) and Kathryn Murphy (Northwestern University) for feedback on the manuscript, and W.\ Callum Wareham (Simon Fraser University) for many fruitful discussions. This work was supported by Natural Sciences and Engineering Research Council of Canada (NSERC) CGS Master's and Doctoral scholarships (J.R.S.), an NSERC Discovery Grant RGPIN-2020-04950 (D.A.S.), and a Tier-II Canada Research Chair CRC-2020-00098 (D.A.S.), and was enabled in part by support provided by the Digital Research Alliance of Canada (alliancecan.ca). 
\end{acknowledgments}

\section*{Data availability}
Code reproducing the results of this article, including \hyperref[fig:fourSpinPanel]{Fig.~\ref*{fig:fourSpinPanel}}, is openly available~\cite{code_ref}. The minimum-work protocols it takes as input are available at Ref.~\cite{sawchuk_2025_15678284}.

\appendix 

\section{Riemannian geometry on the simplex \label{app:geom}}
Essential definitions from Riemannian geometry are collected here for reference; see Refs.~\cite{cheeger_comparison_1975, doCarmoGeometry, berger_panoramic_2003} for complete treatments. On a Riemannian manifold $(M,g)$ with local coordinates $x^i$, a tangent vector $X = X^i \partial_i$ has components $X^i$, the metric has components $g_{ij} = g(\partial_i,\partial_j)$, and (with summation over repeated indices implied) 
\begin{equation} \label{eq:innerProduct}
    g(X,Y) = g_{ij}X^i Y^j \ , \qquad \norm{X}= \sqrt{g(X,X)}\ .
\end{equation}
Following the standard conventions of tensor calculus, in this appendix we write the components of vectors (e.g., a velocity $\dot{\eq}$ or perturbation $\delta \eq$ on $\simp$) with upper, or \textit{contravariant}, indices, and those of covectors (linear functions of vectors, such as the gradient of a scalar) with lower, or \textit{covariant}, indices.

\subsection{Coordinates and conventions on the simplex \label{app:geom:coords}}
Natural coordinates for the thermodynamic manifold $(\simp, L^+)$ are the equilibrium probabilities $\pi^0,\dots,\pi^n$, in which the inner product is the matrix contraction
\begin{equation}
    g_{\eq}(X,Y) = X\t L^+(\eq) \, Y \ .
\end{equation}
These coordinates are redundant due to the normalization of probability ($n+1$ coordinates for an $n$-dimensional manifold), so tangent vectors are the zero-sum vectors $X \in \one^\perp$, and the textbook formulas in this appendix apply with two amendments. 

First, since $L^+$ is singular, the role of the metric inverse $g^{ij}$ is played by its pseudoinverse $L$. This is a true inverse on the tangent space, since
\begin{equation}
    LL^+ = L^+L=I-\tfrac{1}{n+1}\one\one\t    
\end{equation}
is the orthogonal projector onto $\one^\perp$. The metric $L^+$ converts a tangent vector $X$ with contravariant components $X^\mu$ into the covector $g(X,\cdot)$ with covariant components $X_\mu = L^+_{\mu\nu}X^\nu$, and $L$ inverts this. These are the index-lowering and index-raising operations of the standard formalism. The variational derivative $\delta \Wex /\delta \eq$ is a covector field along the protocol, and $L$ raises its index in \hyperref[eq:IIgrad]{Eq.~(\ref*{eq:IIgrad})}, mapping it to a tangent vector, the contravariant gradient.

Second, the metric is only formally defined on $\simp$, but the partial derivatives $\partial_\mu = \partial/\partial \pi^\mu$ in \hyperref[eq:christoffel]{Eq.~(\ref*{eq:christoffel})} vary each coordinate independently and therefore require that we extend $L^+$ to a neighborhood of the simplex in $\mathbb{R}^{n+1}_{>0}$. Individual Christoffel symbols depend on the choice of extension, but the dependence drops out of every geometric quantity computed from them (covariant derivatives, geodesics, and curvatures evaluated on tangent vectors). To see this, let $g^{(1)}(\mathbf{x})$, $g^{(2)}(\mathbf{x})$ be two such extensions defined for $\mathbf{x} \in \mathbb{R}^{n+1}_{>0}\supset \simp$. Examples include $L^+$ and $-\Deq^{-1}\rma^{\mathcal{D}}$ (\hyperref[ssec:manifold]{Sec.~\ref*{ssec:manifold}}), or any of the other equivalent forms of the metric from Ref.~\cite{sawchuk_thermodynamic_2026} evaluated on unnormalized arguments. These are extensions of the same metric on the simplex if for all $\eq \in \simp$ and all $X, Y \in \one^{\perp}$, 
\begin{equation}
    X\t [g^{(1)}(\eq) - g^{(2)}(\eq)  ]Y \equiv X\t \delta g(\eq) Y = 0 \ ,
\end{equation}
i.e., $g^{(1)}$ and $g^{(2)}$ describe the same geometry on the simplex. Since their difference $\delta g(\eq)$ is symmetric and annihilates all vectors in $\one^{\perp}$, it must take the form $\delta g(\eq) = \bm{a}(\eq) \one\t + \one \bm{a}(\eq)\t$ for some vector field $\bm{a}(\eq)$. Off the simplex, with a suitable extension of $\bm{a}(\cdot)$,
\begin{equation}
    \delta g(\mathbf{x}) = \bm{a}(\mathbf{x})\one\t + \one \bm{a}(\mathbf{x})\t + F(\mathbf{x}) \ ,
\end{equation}
for a matrix field $F$ such that $F(\eq) = 0$. In any direction along the simplex, the directional derivative of the matrix field $F$ vanishes: $X^\mu \partial_{\mu} F_{\nu\lambda}(\eq) = 0$ for all $X \in \one^\perp$. Hence $\partial_\mu F_{\nu\lambda}(\eq) = 1_\mu c_{\nu\lambda}(\eq)$ for some function $c_{\nu\lambda}(\eq)$, and
\begin{equation}
    \partial_\mu (\delta g)_{\nu\lambda}(\eq) = (\partial_{\mu} a_\nu) 1_\lambda + 1_\nu (\partial_\mu a_\lambda ) + 1_\mu c_{\nu\lambda} \ .
\end{equation}
This vanishes under contraction with a tangent vector in any index. In particular, the contraction of Christoffel symbols~\eqref{eq:christoffel} with tangent vectors is independent of one's choice of extension. The same argument carries to second derivatives, so contraction of the Riemann curvature tensor~\eqref{eq:RiemannCoords} with tangent vectors [giving, e.g., the sectional curvatures~\eqref{eq:sectional}] is independent of extension. Here, we work with the graph-Laplacian form, so we simply evaluate the conductances $\cond_{\mu\nu}(\pi_\mu,\pi_\nu)$ at unnormalized $\eq$. A Riemannian calculus on the probability simplex in exactly this pseudoinverse Laplacian form is developed rigorously in Ref.~\cite{liTransportInformationGeometry2022} for the linear conductance $\cond_{\mu\nu} = (\pi_\mu + \pi_\nu)/2$.

\subsection{Covariant derivatives, geodesics, and parallel transport \label{app:geom:geodesics}}
The covariant derivative operator determined by the metric is the \textit{Levi-Civita connection} $\nabla$. In the coordinates $\curly{\pi^\mu}$, its components, with all indices down, are the Christoffel symbols
\begin{equation}\label{eq:christoffel}
    \Gamma_{\lambda\mu\nu} = \tfrac{1}{2}\paren{\partial_\mu L^+_{\nu\lambda}+\partial_\nu L^+_{\lambda \mu} - \partial_\lambda L^+_{\mu\nu}} \ ,
\end{equation}
symmetric in its second and third indices ($\Gamma_{\lambda\mu\nu}=\Gamma_{\lambda\nu\mu}$). Then the \textit{covariant derivative} of a vector field $Y$ along another vector field $X$ is 
\begin{equation} \label{eq:covDeriv}
    (\nabla_X Y)^\kappa = X^{\mu}\partial_\mu Y^\kappa + L^{\kappa \lambda} \Gamma_{\lambda \mu \nu}X^\mu Y^{\nu} \ .
\end{equation}
The first term on the right-hand side is the standard coordinate directional derivative and the second term is a geometric correction. The covariant derivative of a vector field $X$ along a curve $\eq(t)$ is $D_tX \equiv \nabla_{\dot{\eq}}X$, and $X$ is \textit{parallel} along the curve if $D_tX =0$. Parallel transport preserves inner products, so an orthonormal frame transported along a curve remains orthonormal; such \textit{parallel frames} are used in \hyperref[sec:comparison]{Sec.~\ref*{sec:comparison}}. 

A \textit{geodesic} is a curve whose velocity is parallel,
\begin{equation} \label{eq:geodesic}
    0 = (D_t \dot{\eq})^\kappa = \ddot{\pi}^\kappa + L^{\kappa \lambda}\Gamma_{\lambda \mu\nu}\dot{\pi}^\mu \dot{\pi}^\nu \ .
\end{equation}
It has constant speed, and the first variation of both its length $\ell$ and energy (here the excess work $\Wex$) vanish. A geodesic is uniquely determined by its initial conditions $(\eq(0),\dot{\eq}(0))$, though multiple geodesics may connect two points, as two meridians between the poles of a sphere, or the long and short arcs of a great circle between two nonantipodal points. 
The coordinate acceleration $\ddot{\eq}$ in \hyperref[eq:geodesic]{Eq.~(\ref*{eq:geodesic})} sums to zero ($\one\t \ddot{\eq}=0$) so a geodesic launched with $\dot{\eq}(0) \in \one^\perp$ remains in the simplex (until, perhaps, it reaches the boundary; \hyperref[app:global]{App.~\ref*{app:global}}).

\subsection{Curvature, the index form, and Jacobi fields \label{app:geom:curvature}}
On a curved manifold, covariant derivatives in two directions do not commute, and the \textit{Riemann curvature tensor}
\begin{equation} \label{eq:riemann}
    R(X,Y)Z \equiv \nabla_X\nabla_Y Z - \nabla_Y \nabla_X Z - \nabla_{[X,Y]}Z 
\end{equation} 
measures their failure to commute, with $[X,Y] \equiv X^\mu\partial_\mu Y - Y^\mu \partial_\mu X$ the commutator of the vector fields. Taking the inner product with a fourth vector gives the scalar $R(X,Y,Z,W) = g\paren{R(X,Y)Z, W}$, which is antisymmetric in $(X,Y)$ and in $(Z,W)$, and symmetric under exchange of the two pairs, $R(X,Y,Z,W)=R(Z,W,X,Y)$. In the coordinates $\curly{\pi^\mu}$, $R(X,Y,Z,W)= R_{\lambda\rho\mu\nu} W^\lambda Z^\rho X^\mu Y^\nu$ with 
\begin{equation} \label{eq:RiemannCoords}
R_{\lambda\rho\mu\nu} = \partial_\mu\Gamma_{\lambda\nu\rho} - \partial_\nu\Gamma_{\lambda\mu\rho} + L^{\kappa\sigma}\paren{\Gamma_{\kappa\nu\lambda}\Gamma_{\sigma\mu\rho} - \Gamma_{\kappa\mu\lambda}\Gamma_{\sigma\nu\rho}} \ .    
\end{equation}
As with the Christoffel symbols, individual components depend on the extension of $L^+$ off the simplex, but their contraction with four tangent vectors does not.

The \textit{sectional curvature} of the tangent plane $\varsigma = \Span\curly{X,Y}$ is
\begin{equation} \label{eq:sectional}
    K(\varsigma) = \frac{R(X,Y,Y,X)}{\norm{X}^2\norm{Y}^2 - g(X,Y)^2} \ .
\end{equation}
The denominator is the squared area of the parallelogram spanned by $X$ and $Y$, so $K$ depends only on the plane $\varsigma$ and not on the specific spanning vectors. Geometrically, $K(\varsigma)$ is the Gaussian curvature of the two-dimensional surface swept out by the geodesics initiated at $\eq$ with initial velocities in $\varsigma$. $K(\varsigma)$ is positive when neighboring geodesics in that surface converge, as on a sphere, and negative when they diverge, as on a saddle-shaped surface.

For vector fields $X,Y$ along a geodesic $\eq^\star$, both vanishing at the endpoints, the \textit{index form} is
\begin{equation} \label{eq:indexFormGen}
    I(X,Y) \equiv \int_0^1 \md t \sbrac{\, g\paren{D_t X, D_t Y} - R\paren{X, \dot{\eq}^\star, \dot{\eq}^\star, Y} \,} \ ,
\end{equation}
whose quadratic form $I(X,X)$ appears in the main text as \hyperref[eq:indexForm]{Eq.~(\ref*{eq:indexForm})}. For a one-parameter family $\eq(t;\epsilon)$ of protocols with $\eq(\,\cdot\,;0)=\eq^\star$, with variation field $X$, the excess work satisfies $\partial_{\epsilon} \Wex\vert_{\epsilon = 0}=0$ and $\partial^2_{\epsilon}\Wex\vert_{\epsilon = 0} = 2 I(X,X)$~\cite{cheeger_comparison_1975}, which gives the expansion of the excess work to second order [\hyperref[eq:secondVariation]{Eq.~(\ref*{eq:secondVariation})}]. The Euler--Lagrange equation of $I$ is the \textit{Jacobi equation}
\begin{equation} \label{eq:jacobi}
    D_t^2 X + R\paren{X, \dot{\eq}^\star}\dot{\eq}^\star = 0 \ .
\end{equation}
This is also the linearization of the geodesic equation~\eqref{eq:geodesic} about $\eq^\star$, so its solutions, the \textit{Jacobi fields}, are the variation fields of one-parameter families of geodesics through $\eq^\star$ whose endpoints need not be fixed.

\subsection{Submanifolds and the second fundamental form \label{app:geom:sff}}
For a submanifold $M \subset M'$ with the inherited metric, the ambient tangent space $T_{\eq}M'$ splits orthogonally into the tangent space $T_{\eq}M$ of the submanifold and its normal space $N_{\eq}M$ with respect to $M'$: $T_{\eq} M' = T_{\eq} M \oplus N_{\eq} M$. This defines tangential and normal projections $X = X^{\parallel} + X^{\perp}$ of any vector field $X$. The connections $\nabla$ of $M$ and $\nabla'$ of $M'$ are related by the \textit{Gauss formula}
\begin{equation} \label{eq:gauss}
    \nabla'_X Y = \nabla_X Y + \II(X,Y) \, , \qquad \II(X,Y)\equiv(\nabla'_X Y)^{\perp} \, , 
\end{equation}
whose normal part $\II$ is the \textit{second fundamental form}, at each point $\eq$ a symmetric bilinear map $T_{\eq}M \times T_{\eq}M \to N_{\eq}M$. In coordinates $u^i$ on $M$ and $v^a$ on $M'$, this reads
\begin{equation} \label{eq:SFFcoords}
    \II^a{}_{ij} = \partial_i \partial_j v^a + \Gamma'\mkern1mu ^a{}_{bc}\,\partial_i v^b \partial_j v^c - \Gamma^k{}_{ij}\,\partial_k v^a \ ,
\end{equation} 
with $\partial_i =\partial/\partial u^i$, and $\Gamma'\mkern1mu^a{}_{bc}$ and $\Gamma^k{}_{ij}$ the respective Christoffel symbols of $M'$ and $M$ in these coordinates. For $M' = \simp$, $v^a$ are the probabilities $\pi^\mu$ and $\Gamma'\mkern1mu^{\mu}{}_{\kappa \lambda} = L^{\mu\nu}\Gamma_{\nu\kappa \lambda}$. Evaluated on the velocity of a curve in $M$, $\II(\dot{\eq},\dot{\eq})$ is the normal component $(D_t \dot{\eq})^\perp$ of the curve's ambient acceleration, analogous in classical mechanics to the acceleration due to a normal force used to enforce some constraint. If $\II = 0$ identically, then $M$ is \textit{totally geodesic} in $M'$, i.e., every geodesic of $M$ is a geodesic of $M'$. The covariant acceleration along a geodesic $\eq^\star_M$ of $M$ is purely normal in the ambient space, so for $M' = \simp$, $\nabla'_{\dot{\eq}^\star_M}\dot{\eq}^\star_M = \II(\dot{\eq}^\star_M,\dot{\eq}^\star_M)$. For smooth variations, comparing~\eqref{eq:covDeriv} to the Euler--Lagrange equation for $\Wex$ shows that $L^+_{\lambda \kappa }(\nabla'_{\dot{\eq}} \dot{\eq})^{\kappa} = - \tfrac{1}{2} \,\delta \Wex / \delta \pi^\lambda$, from which the relation between $\II$ and the gradient of the excess work [\hyperref[eq:IIgrad]{Eq.~(\ref*{eq:IIgrad})}] follows directly.

\subsection{Isometries \label{app:geom:iso}}
Let $\varphi$ be a smooth invertible map of a manifold onto itself. If $X$ is a vector field taking the value $X\vert_{\bm{\pi}}$ at a point $\bm{\pi}$, its \textit{pushforward} under $\varphi$ is the vector field 
\begin{equation} \label{eq:pushforward}
    \varphi_* X \big\vert_{\varphi(\bm{\pi})} = \mathbb{J}_{\varphi}X\big\vert_{\bm{\pi}} 
\end{equation}
with $\mathbb{J}_{\varphi} \equiv \partial \varphi/\partial\eq$ the Jacobian matrix of $\varphi$. That is, $\varphi_*$ transforms both the vectors and the points on the manifold to which they are attached. The map $\varphi$ is said to be an \textit{isometry} of the manifold if it preserves the metric, i.e., if the inner product of any two tangent vectors is unchanged when both are pushed forward by $\varphi$. In terms of the metric $L^+$, this reads 
\begin{equation} \label{eq:isometry}
    \mathbb{J}_{\varphi}\t\,  L^+\!\paren{\varphi(\eq)}\,\mathbb{J}_{\varphi} = L^+(\eq) \ ,\quad \forall \, \eq\in \simp\ .
\end{equation}
An isometry preserves the length and energy of every curve, and therefore maps geodesics to geodesics. It also preserves the Levi-Civita connection~\eqref{eq:covDeriv}, and hence the curvature tensor~\eqref{eq:riemann}:
\begin{subequations} \label{eq:connectionCommutation}
    \begin{align}
        \nabla_{(\varphi_* X)}\,\varphi_* Y &= \varphi_*(\nabla _X Y) \label{eq:isoConn} \ , \\
         R(\varphi_* X, \varphi_* Y) \varphi_*Z &= \varphi_*\big(R(X,Y)Z \big) \label{eq:isoCurv} \ .
    \end{align}
\end{subequations}
Since the covariant derivative along a curve $\bm{\pi}(t)$ is $\nabla$ in the direction of a curve's velocity,~\eqref{eq:isoConn} means that $D_t(\mathbb{J}_{\varphi} X)$ evaluated along the transformed curve $\varphi(\bm{\pi}(t))$ is equal to $\mathbb{J}_{\varphi}D_t X$ evaluated along $\bm{\pi}(t)$.

\section{Dynamical symmetries \label{app:symmetry}}
This appendix collects the definitions, proofs, and coarse-graining construction behind \hyperref[sec:sym]{Sec.~\ref*{sec:sym}}.

\subsection{Dynamical symmetries are isometries of the thermodynamic manifold \label{app:symproofs}}
We show that a permutation of the states is a dynamical symmetry if and only if it acts isometrically on the thermodynamic manifold [\hyperref[eq:R1]{Eq.~(\ref*{eq:R1})}]. A permutation $\sigma$ is a dynamical symmetry when the transition rates are equivariant, 
\begin{equation} \label{eq:rateEquivariance}
    \rma(\eq) = P_{\sigma}\t \, \rma(P_{\sigma}\eq) \, P_{\sigma} \ .
\end{equation}
In components, this reads $w_{\sigma(\mu)\sigma(\nu)}(P_{\sigma}\eq) = w_{\mu\nu}(\eq)$, so $\sigma$ carries each transition to a transition with the same rate law. We identify the group of such permutations with the automorphism group $\Aut(G)$ in \hyperref[app:colored]{App.~\ref*{app:colored}}. Since $D_{P_{\sigma}\eq} = P_{\sigma} D_{\eq} P_{\sigma}\t$ and $D_{\eq}$ is invertible, the equivariance of the rates holds if and only if the Laplacian $L = - \rma D_{\eq}$ satisfies the same equivariance [\hyperref[eq:isocriterion]{Eq.~(\ref*{eq:isocriterion})}]. Since $(U\t A U)^+ = U\t A^+ U$ for any orthogonal $U$, and since pseudoinversion is an involution [$(A^+)^+ = A$], equivariance of $L$ is equivalent to equivariance of $L^+$:
\begin{equation} \label{eq:isoLplus}
    P_\sigma\t\, L^+\!\paren{P_\sigma \eq}\, P_\sigma = L^+(\eq) \ , \qquad \forall\, \eq \in \simp \ ,
\end{equation}
which is the isometry condition~\eqref{eq:isometry} with $\mathbb{J}_{P_{\sigma}} = \partial (P_{\sigma} \eq) / \partial \eq = P_{\sigma}$. Each step is an equivalence, so the permutations acting isometrically on the thermodynamic manifold are exactly the dynamical symmetries. Note that the isometry group may contain maps that are not permutations of the states (e.g., continuous symmetries), so this does not imply that $\Aut(G) = \Iso(\simp)$. 

Now let $\sigma \in \Aut(G,M')$ for a submanifold $M' \subseteq \simp$ [\hyperref[eq:stabilizerGroup]{Eq.~(\ref*{eq:stabilizerGroup})}]. Since $P_{\sigma}$ maps $M'$ onto $M'$, for all $X, Y \in T_{\eq} M'$
we have
\begin{equation}
    (P_{\sigma}X)\t L^+ (P_{\sigma}\eq)\, (P_{\sigma}Y) = X\t \, L^+(\eq) \, Y \ ,
\end{equation}
and thus $P_{\sigma}$ is also an isometry of $M'$. For $H \subseteq \Aut(G,M')$, the fixed-point set of $H$ in $M'$ is $M_H \cap M'$, and each of its connected components is totally geodesic in $M'$~\cite{kobayashi_geomtry, berger_panoramic_2003}.

\subsection{Dynamical symmetries are automorphisms of the Markov graph \label{app:colored}}
We show that the group of dynamical symmetries for a system can be identified with a suitably defined automorphism group for its Markov graph. By locality and detailed balance (\hyperref[ssec:setup]{Sec.~\ref*{ssec:setup}}), the transition rates obey 
\begin{equation} \label{eq:wy}
    w_{\mu\nu}(y) = y\, w_{\nu\mu}(y^{-1})\,, \qquad y \equiv \pi_{\mu}/\pi_\nu  \ .
\end{equation}
Assign to each ordered pair of states a \textit{color} $\mathcal{C}(\mu,\nu) \equiv w_{\mu\nu}(\cdot)$ associated with its rate law, so that two transitions share a color if and only if their rates are the same function of $y$. The colors may identify a shared physical process, e.g., two edges corresponding to flips of kinetically indistinguishable spins have the same color. 

Under $P_{\sigma}$, $\pi_{\sigma(\mu)} / \pi_{\sigma(\nu)} \mapsto \pi_{\mu}/\pi_{\nu}$, so the equivariance~\eqref{eq:rateEquivariance} reads $w_{\sigma(\mu) \sigma(\nu)}(\cdot) = w_{\mu\nu}(\cdot)$, which holds if and only if $\sigma$ preserves the colors. The dynamical symmetries therefore form the group
\begin{equation} \label{eq:coloredAut}
    \Aut(G) \equiv \curly{\sigma \,: \, \mathcal{C}(\sigma(\mu),\sigma(\nu)) = \mathcal{C}(\mu,\nu) \ \ \forall \ \mu\neq\nu \in \Omega}\, ,
\end{equation}
the automorphism group of the Markov graph with colored edges. Pairs of states not joined by an edge have $w_{\mu\nu}\!=\!0$, so every $\sigma\!\in\!\Aut(G)$ carries edges to edges, preserving the adjacency of the graph. 

If all transitions obey a common rate law $w(y)$, as for Glauber dynamics $w(y) = w_0/[1+y^{-1}]$ at a common bare rate $w_0$, then every edge has the same color and $\Aut(G)$ is the ordinary automorphism group of the Markov graph. This is the case for the four-spin system of the main text. 

\begin{figure}[b]
    \centering
    \includegraphics[]{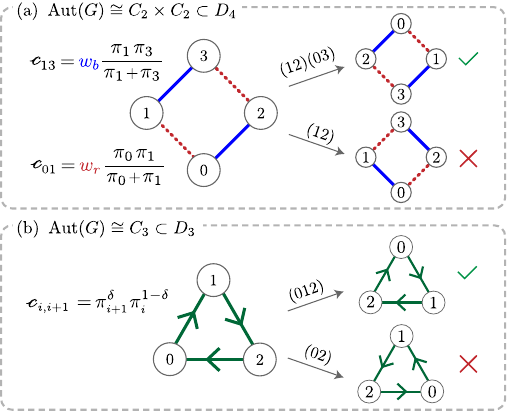}
    \caption{A permutation of the states is a dynamical symmetry (green check marks) only if every edge of the Markov graph keeps its color and orientation. (a) Two spins with bare flip rates $w_b$ (solid blue edges) and $w_r$ (dotted red edges). (b) Three-state ring with splitting factor $\delta \neq \frac{1}{2}$ and common bare rate $w_0 = 1$, so each edge has the same color but a fixed direction aligned with the transitions obeying $w(y) = y^{\delta}$, with $y$ as in~\eqref{eq:wy}.}
    \label{fig:symExamples}
\end{figure}

If the rate laws differ between edges, but each is the same in both directions [$\mathcal{C}(\mu,\nu) = \mathcal{C}(\nu,\mu)$], then the conductance $\cond_{\mu\nu}$ is a symmetric function of $\pi_\mu$ and $\pi_\nu$, and the color is assigned to the undirected edge. $\Aut(G)$ is then the automorphism group of the \textit{edge-colored Markov graph}. For example, the Markov graph for two spins with states $0 = (\downarrow\downarrow)$, $1=(\uparrow\downarrow)$, $2=(\downarrow\uparrow)$, and $3=(\uparrow\uparrow)$ is the four-cycle $0$--$1$--$3$--$2$--$0$. If the spins flip at different bare rates [$w_b$ and $w_r$ in \hyperref[fig:symExamples]{Fig.~\ref*{fig:symExamples}(a)}], the colors alternate around the cycle and $\Aut(G) = \curly{\mathrm{id}, (01)(23),(02)(13),(03)(12)} \cong C_2 \times C_2$, half the group $D_4$ obtained for equal bare rates. [In cyclic notation, $(\mu_0 \mu_1\cdots \mu_{a-1})$ is the permutation $\mu_{j} \to \mu_{j +1 \mod{a}}$, so $(12)(03)$ sends $\curly{0,1,2,3} \to \curly{3,2,1,0}$].

If the rate law depends on the direction an edge is traversed, the conductance $\cond_{\mu\nu}$ is not a symmetric function of $\pi_\mu$ and $\pi_\nu$, and $\mathcal{C}(\nu,\mu) \neq \mathcal{C}(\mu,\nu)$. We may then represent each transition by a directed edge to define the \textit{edge-colored Markov digraph}, and $\Aut(G)$ is its automorphism group. Each edge carries only one independent color under detailed balance, so it is sufficient to choose a color and a reference direction for each unique pair of rates. For example, consider a simple three-state model of a rotary motor in the absence of any chemical driving force: The clockwise transitions are $w_{i+1,i}(y) = w_0\, y^{\delta}$, and the counterclockwise transitions are $w_{i,i+1}(y) = w_0 \,y^{1-\delta}$, with $w_0$ the bare rate and $\delta \in [0,1]$ the splitting factor~\cite{brownAllocatingSplittingFree2018, brownTheoryNonequilibriumFree2020}. For $\delta = \tfrac{1}{2}$, the edges are undirected and $\Aut(G) \cong D_3$ (the symmetries of the equilateral triangle in the plane), but if $\delta \neq \tfrac{1}{2}$, the reflections reverse every directed edge, leaving $\Aut(G) \cong C_3$ [\hyperref[fig:symExamples]{Fig.~\ref*{fig:symExamples}(b)}].

\subsection{Exact excess work \label{ssec:exactExcess}}
Here we show that the stationarity of symmetric protocols and the decoupling of the second variation hold for arbitrary driving speed. Suppose that $\bm{p}(t)$ is the solution of the master equation $\dot{\bm{p}}(t) = \rma(\eq(t))\, \bm{p}(t)$ with rates evolving via the control protocol $\eq(t)$, and initial condition $\bm{p}(0) = \eq_0$ satisfying $P_{\sigma}\,\eq_0 = \eq_0$. Then it follows from the equivariance of the rates~\eqref{eq:rateEquivariance} that $P_{\sigma} \, \bm{p}(t)$ is the solution of the master equation with rates evolving via the control protocol $P_{\sigma}\,\eq(t)$, from the same initial condition. 

Let $H \subseteq \Aut(G,M')$ fix protocol endpoints $\eq_0, \eq_1 \in M'$. Then for any $\sigma \in H$, by orthogonality of $P_{\sigma}$ and the fact that $P_{\sigma}\,\eq(\bm{V}) = \eq(P_{\sigma}\bm{V})$,
\begin{subequations} \label{eq:exactInvariance}
\begin{align}
    \langle \mathcal{W}_{\rm ex}\rangle [P_{\sigma}\eq] &= \kB T\int_0^\tau\md t\, (P_{\sigma}\,\bm{p}-P_{\sigma}\,\eq)\t P_{\sigma}\dot{\bm{V}} \\
    &= \langle \mathcal{W}_{\rm ex} \rangle[\eq] \ .
\end{align}
\end{subequations}
Thus the exact excess work is invariant under graph automorphisms consistent with the symmetry of the protocol endpoints.

For $M = M_H \cap M'$, consider a protocol $\eq^\star \in \mathcal{P}_{M}$, which is stationary in the excess work among $H$-invariant protocols. This means that the first variation of the excess work vanishes for any variation field $X^{\parallel} \in T_{\eq^\star}\mathcal{P}_{M}$ tangent to the symmetric subspace, 
\begin{equation} \label{eq:firstVarVanish}
    \delta \! \angbr{\mathcal{W}_{\rm ex}}[\eq^\star;X^{\parallel}] = \frac{\md}{\md \epsilon}\bigg\vert_{\epsilon=0}\angbr{\mathcal{W}_{\rm ex}}[\eq^\star + \epsilon X^{\parallel}]=0 \ .
\end{equation}
It remains to show that the first variation of the excess work at such a protocol vanishes with respect to \textit{any} variation field $X \in T_{\eq^\star}\mathcal{P}_{M'}$, including those that break the symmetries of the protocol. Since $\eq^\star$ is a symmetric protocol, $\eq^\star + \epsilon P_{\sigma} X = P_{\sigma}(\eq^\star + \epsilon X)$, and so by the $H$-invariance of the excess work~\eqref{eq:exactInvariance}, the first variation of the excess work is identical for perturbations $X$ and $P_{\sigma}X$. This is true for any $\sigma \in H$, so by linearity of the first variation,
\begin{subequations}
    \begin{align}
        \delta\langle \mathcal{W}_{\rm ex}\rangle[\eq^\star;X] &= \frac{1}{|H|}\sum_{\sigma \in H}\delta\langle \mathcal{W}_{\rm ex}\rangle [\eq^\star; P_{\sigma} X] \\
        &= \delta\langle\mathcal{W}_{\rm ex}\rangle [\eq^\star; \Pi_0 X]  \\
        &= 0 \, .
    \end{align}
\end{subequations}
This follows from stationarity in $M$~\eqref{eq:firstVarVanish} and the fact that $\Pi_0$ is the projector onto the tangent space of $M$. Thus $\eq^\star$ is a stationary point of the exact excess work on $\mathcal{P}_{M'}$.

A similar argument applies at second order. Define the second variation at $\eq^\star$ by
\begin{equation}
    \delta^2\langle \mathcal{W}_{\rm ex}\rangle[X,Y] \equiv \frac{\md}{\md \epsilon}\Bigg\vert_{\epsilon=0}\frac{\md}{\md \eta}\Bigg\vert_{\eta=0}\langle \mathcal{W}_{\rm ex}\rangle[\eq^\star + \epsilon X + \eta Y] \ .
\end{equation}
It immediately follows from~\eqref{eq:exactInvariance} that
\begin{equation}
   \delta^2\langle \mathcal{W}_{\rm ex}\rangle[P_{\sigma}X,Y] = \delta^2\langle \mathcal{W}_{\rm ex}\rangle[X,P_{\sigma}\t Y] \ .
\end{equation}
Let $X_{\rho} \equiv \Pi_{\rho} X$ be the part of a perturbation $X$ in sector $\mathcal{U}_{\rho}$, with $\Pi_{\rho}$ the sector projectors of \hyperref[app:sectors]{App.~\ref*{app:sectors}}. These are symmetric and mutually orthogonal~\eqref{eq:Piprops}, and since they are real linear combinations of the $P_{\sigma}$~\eqref{eq:sectorProjector}, the identity above holds with $\Pi_{\rho}$ in place of $P_{\sigma}$. Thus for any $\rho \neq \rho'$,
\begin{subequations}
    \begin{align}
        \delta^2\langle \mathcal{W}_{\rm ex}\rangle[X_{\rho},Y_{\rho'}] &= \delta^2\langle \mathcal{W}_{\rm ex}\rangle[X,\Pi_{\rho}\Pi_{\rho'}Y] \\
        &= \delta^2\langle \mathcal{W}_{\rm ex}\rangle[X,0] \\
        &= 0 \ .
    \end{align}
\end{subequations}
The sector decomposition of the second variation along symmetric optima therefore holds at any driving speed, though the curvature form of~\eqref{eq:R2index} is valid only for slow driving.

\subsection{Symmetry and coarse-graining \label{app:CG}}
Here we show that symmetric control is exactly full control of a coarse-grained model. For subgroup $H \subseteq \Aut(G)$ with $r$ orbits $\mathcal{O}_1,\dots,\mathcal{O}_r$, define the \textit{lumping map}~\cite{hummer_optimal_2015} 
\begin{equation} \label{eq:lump}
    \Lambda\in\mathbb{R}^{r\times(n+1)}, \ \ \Lambda_{\mathcal{O}\mu}= \begin{cases}
        1 & \mu \in \mathcal{O}, \\
        0 & \text{otherwise.} 
    \end{cases} 
\end{equation}
The lumping map acts as a bijection between $\eq \in M_H$ and the normalized \textit{orbit masses} $\hat{\eq} \equiv \Lambda \eq \in \Delta^{r-1}$, so the orbit masses are coordinates on the symmetric submanifold $M_H$. The pseudoinverse $\Lambda^+ =\Lambda\t(\Lambda\Lambda\t)^{-1}$ spreads each orbit mass uniformly over its orbit, $(\Lambda^+ \hat{\eq})_{\mu} = \hat{\pi}_{\mathcal{O}}/|\mathcal{O}|$. The product 
$\Lambda^+ \Lambda = \Pi_0$~\eqref{eq:Pi0} is the projector onto the orbit-uniform vectors, so it acts as the identity on $M_H$. This establishes that there exists a one-to-one map between points on $M_H$ and points on $\Delta^{r-1}$.

It remains to show that the induced metric on $\Delta^{r-1}$ is the pseudoinverse of the weighted graph Laplacian of the lumped chain. For some driving velocity $\dot{\eq} \in T_{\eq}M_H$, the instantaneous excess power is $\dot{\eq}\t L^+ \dot{\eq} = \bm{\phi}\t \dot{\eq}$ with $\bm{\phi} = L^+ \dot{\eq}$ the conjugate potential (\hyperref[sec:convexity]{Sec.~\ref*{sec:convexity}}). The tangent vector $\dot{\eq}$ is $H$-invariant by definition, and $L^+$ commutes with $P_{\sigma}$ at $\eq \in M_H$, so $\bm{\phi}$ is also $H$-invariant. The conjugate potential $\hat{\bm{\phi}}$ of the lumped chain for which the excess power is invariant ($\bm{\phi}\t \dot{\eq} = \hat{\bm{\phi}}\t\dot{\hat{\eq}}$) is given by the transformation $\bm{\phi} = \Lambda\t \hat{\bm{\phi}}$, thus
\begin{subequations}
    \begin{align}
        \dot{\hat{\eq}} &= \Lambda L \bm{\phi} \\ 
        &= \Lambda L \Lambda\t \hat{\bm{\phi}} \\
        &\equiv \hat{L} \hat{\bm{\phi}}
    \end{align}
\end{subequations}
and therefore $\hat{\bm{\phi}} = \hat{L}^+\dot{\hat{\eq}}$ (up to an additive constant), so
\begin{equation}
    \dot{\eq}\t L^+ \dot{\eq} = \dot{\hat{\eq}}\t \hat{L}^+ \dot{\hat{\eq}} \ ,
\end{equation}
and thus the metric induced on $\Delta^{r-1}$ is $\hat{L}^+ = (\Lambda L \Lambda\t)^+$. In the electrical-network picture, the nodes of an orbit sit at a common potential, so edges within an orbit carry no current and each orbit can be merged into a single node. The edges joining two orbits then act as resistors in parallel, with conductances adding as $\hat{\cond}_{\mathcal{O}\mathcal{O}'} = \sum_{\mu\in \mathcal{O}}\sum_{\nu \in \mathcal{O}'} \cond_{\mu\nu} = -\hat{L}_{\mathcal{O}\mathcal{O}'}$.

We now show that $\hat{L}$ is the weighted Laplacian of the lumped chain in the sense of \hyperref[ssec:manifold]{Sec.~\ref*{ssec:manifold}}. At a point $\eq \in M_H$, the total rate into an orbit $\mathcal{O}$ from a state $\nu$ depends only on the orbit of $\nu$, following directly from the equivariance of the rates~\eqref{eq:rateEquivariance},
\begin{equation}
    \sum_{\mu\in \mathcal{O}}w_{\mu\, \sigma(\nu)} = \sum_{\mu \in \mathcal{O}}w_{\sigma^{-1}(\mu)\,\nu} = \sum_{\mu\in \mathcal{O}}w_{\mu\nu} \ ,
\end{equation}
for all $\sigma \in H$. The transition rate from $\mathcal{O}'$ to $\mathcal{O}$,
\begin{equation} \label{eq:lumpability}
    \hat{w}_{\mathcal{O}\mathcal{O}'} = \sum_{\mu \in \mathcal{O}}w_{\mu\nu} \ \ \text{for any } \nu \in \mathcal{O}' \ ,
\end{equation}
is therefore well defined. In matrix form, \hyperref[eq:lumpability]{Eq.~(\ref*{eq:lumpability})} is $\hat{\rma}\Lambda = \Lambda \rma$, the \textit{lumpability} condition of Kemeny and Snell~\cite{kemenyFiniteMarkovChains1983, tianLumpabilityCommutativityMarkov2006}. Applying $\Lambda$ to the master equation under a symmetric protocol then gives a closed master equation for the orbit masses with rate matrix $\hat{\rma}$. Generally, coarse-grained thermodynamics keeps the structure of the microscopic theory when the states within each mesostate remain equilibrated among themselves, usually satisfied only approximately by a separation of timescales~\cite{espositoStochasticThermodynamicsCoarsegraining2012}. Here this condition holds at any driving speed, as any distribution uniform within each orbit remains uniform under a symmetric protocol~\footnote{The condition $\hat{\rma}\Lambda = \Lambda \rma$ is equivalent to the commutation of the rate matrix $\rma(\eq(t))$ with the projector $\Pi_0 = \Lambda^+ \Lambda$ at $\eq(t) \in M_H$, from which $(I - \Pi_0)\,\rma(\eq(t)) \,\Pi_0 = 0$. Then if $\bm{p}(t)$ is orbit-uniform, its velocity $\dot{\bm{p}}(t) = \rma(\eq(t))\bm{p}(t)$ is also orbit-uniform, hence the dynamics preserve orbit-uniformity.}. The weighted Laplacian of the lumped chain is
\begin{subequations}
    \begin{align}
        \hat{L} &= -\hat{\rma}D_{\hat{\eq}} \\
        &= - \hat{\rma}\Lambda D_{\eq}\Lambda\t \\
        &= \Lambda(-\rma D_{\eq})\Lambda\t \ .
    \end{align}
\end{subequations}
Thus $(\Delta^{r-1} ,\hat{L}^+)$ is a thermodynamic manifold and $\Lambda$ is an isometry from $(M_H, L^+)$ onto it, so lengths, geodesics, and optimal costs all coincide. Symmetric control is therefore exactly full control of the lumped chain.

\section{Symmetry sectors and the second variation \label{app:secondVar}}
Throughout this appendix, $H \subseteq \Aut(G,M')$, $M = M_H \cap M'$ is connected, and $\eq^\star_M \in \mathcal{P}_{M}$ is a minimum-work protocol, hence a geodesic of both $M$ and $M'$ with $\norm{\dot{\eq}^\star_M}^2 = \Wex^\star_M$.

\subsection{Symmetry sectors and real irreps \label{app:sectors}}
Here we provide the representation theory behind \hyperref[sec:comparison]{Sec.~\ref*{sec:comparison}}. Everything here is standard, so we do not reproduce the proofs; see Refs.~\cite{serreLinearRepresentationsFinite1977, chenGroupRepresentationTheory2002, cottonChemicalApplicationsGroup1990} for complete treatments. 

A \textit{representation} of a group $H$ assigns to each $\sigma \in H$ an invertible matrix such that the multiplication table of the matrices agrees with the multiplication table of the group. The permutation matrices $P_{\sigma}$ represent $H$ on the space of state vectors $\mathbb{R}^{n+1}$ (with $|\Omega|=n+1$), and the action of $\sigma$ on $m$ control parameters defines another representation, with matrices $S_{\sigma} \in \mathbb{R}^{m\times m}$. 

Given a representation of a finite group on a real vector space, it is always possible to decompose the space as a direct sum of minimal $H$-invariant subspaces (subspaces closed under every element of $H$). These are the real irreducible representations, or \textit{irreps}. Two minimal $H$-invariant subspaces are considered to be copies of the same irrep when they transform identically under $H$, i.e., if their bases can be chosen so that the matrices representing $H$ on the subspaces coincide. Label the distinct real irreps of $H$ as $\rho =0,1,2,\dots$, each with dimension $d_{\rho}$. A representation may contain several copies of the same irrep, and any mixture of those copies gives another valid decomposition, so the splitting into real irreps is not unique. This ambiguity is removed by grouping the copies of $\rho$ into \textit{sectors} (or \textit{isotypic components}). Writing $U_{\rho}$ for the sector carrying irrep $\rho$ in the permutation representation, the unique decomposition is
\begin{equation} \label{eq:isotypicDecomposition}
    \mathbb{R}^{n+1} = \bigoplus_{\rho} U_{\rho} \, , \qquad n+1 = \sum_{\rho}m_{\rho}d_{\rho} 
\end{equation}
Here $m_{\rho}$ is the number of copies of $\rho$ (a property of the representation, akin to the degeneracy of an eigenvalue), so $\dim U_{\rho} = m_{\rho} d_{\rho}$. We reserve $\rho = 0$ for the \textit{trivial representation} in all cases, which assigns the identity matrix to each element of $H$, so the corresponding \textit{symmetric sector} $U_0$ is the set of directions in $\mathbb{R}^{n+1}$ unchanged by any $P_{\sigma}$. This sector is spanned by the orbit indicators $(\bm{1}_{\mathcal{O}})_{\mu} = 1$ for $\mu \in \mathcal{O}$ and $0$ otherwise, so $(d_0, m_0) = (1,r)$ with $r$ the number of orbits in the group.

Let $\Pi_{\rho}$ be the orthogonal projector onto $U_{\rho}$. In particular, $\Pi_0$ is the orbit average of~\eqref{eq:Pi0}. These satisfy 
\begin{subequations} \label{eq:Piprops}
    \begin{align}
        \Pi_{\rho}\t &= \Pi_{\rho} \, , \\
        \Pi_{\rho}\Pi_{\rho'} &= \delta_{\rho \rho'} \Pi_{\rho} \, , \\
        \sum_{\rho} \Pi_{\rho} &= I \, .
    \end{align}
\end{subequations}
We construct these projectors below in~\hyperref[eq:sectorProjector]{Eq.~(\ref*{eq:sectorProjector})}, and we will see that each is a real linear combination of the $P_{\sigma}$. Then if $B$ is any operator that commutes with every $P_{\sigma}$ [e.g., the metric $L^+(\eq)$ or rate matrix $\rma(\eq)$ at a symmetric distribution $\eq \in M_H$, or the curvature operator along a symmetric protocol~\eqref{eq:permComm}], $B$ also commutes with the projectors $\Pi_{\rho}$. Thus from~\eqref{eq:Piprops}, $\Pi_{\rho} B\, \Pi_{\rho'} = 0$ for $\rho \neq \rho'$, so 
\begin{equation} \label{eq:simpleDecomp}
    B = \sum_{\rho}\Pi_{\rho} \,B \, \Pi_{\rho} \, ,
\end{equation}
hence $B$ is block-diagonal in a basis adapted to the sectors. This is all that is required for the sector decomposition of the second variation in \hyperref[app:comparison]{App.~\ref*{app:comparison}}.

Each block has further structure, because $U_{\rho}$ contains $m_{\rho}$ copies of one irrep. In a suitable basis~\cite{murotaNumericalAlgorithmBlockdiagonal2010a},
\begin{equation} \label{eq:equivariantForm}
    B = \bigoplus_{\rho}B_{\rho} =  \bigoplus_{\rho}\big(I_{d_{\rho}/e_{\rho}} \otimes A_{\rho}\big)\, ,
\end{equation}
i.e., the restriction $B_{\rho}$ of $B$ to $U_{\rho}$ is itself block diagonal, comprising $d_{\rho}/e_{\rho}$ identical copies of an $(e_{\rho}m_{\rho}) \times (e_{\rho}m_{\rho})$ matrix $A_{\rho}$. Here $e_{\rho}$ takes one of three values, depending on the type of irrep: $e_{\rho} = 1$ for real-type irreps, $e_{\rho} =2$ for complex-type irreps, and $e_{\rho}=4$ for quaternionic-type irreps~\cite{serreLinearRepresentationsFinite1977}. Complex-type irreps split into two complex conjugate irreps. Consider the cyclic group $C_3$ of \hyperref[fig:symExamples]{Fig.~\ref*{fig:symExamples}(b)}, for example. Its two-dimensional real irrep is the rotation of a plane by $2\uppi/3$, which splits into inequivalent one-dimensional complex irreps generated by $\e{2 \uppi \,i/ 3}$ and $\e{-2 \uppi \,i/ 3}$. Quaternionic-type irreps split into two copies of one complex irrep. Every group in the main text has only real-type irreps, in which case $B_{\rho} = I_{d_{\rho}} \otimes A_{\rho}$ with $A_{\rho}$ an arbitrary $m_{\rho} \times m_{\rho}$ matrix. In particular, $B_{\rho}$ is a multiple of the identity when $m_{\rho} = 1$. 

In practice, none of the irreps need to be constructed explicitly. To decompose the representation, one only needs the \textit{character} $\chi_{\rho}(\sigma)$ of each real irrep, the trace of the real $d_{\rho}\times d_{\rho}$ matrix representing $\sigma$ on a single copy of $\rho$ [in particular, $d_{\rho} = \chi_{\rho}(\rm id)$ since the identity element $\rm id$ is always represented by an identity matrix]. \textit{Character tables} for finite groups, which list the characters of the complex irreps, are available in standard references~\cite{cottonChemicalApplicationsGroup1990, conwayAtlasFiniteGroups1985} and can be computed algorithmically~\cite{GAP4}. Each complex irrep can be identified as being of real, complex, or quaternionic type by a simple calculation on its character~\cite[Prop.~39]{serreLinearRepresentationsFinite1977}, and each real character $\chi_{\rho}$ is then obtained from the complex characters. 
In terms of the real characters,   
\begin{equation} \label{eq:irrepType}
    e_{\rho} = \frac{1}{|H|} \sum_{\sigma \in H}[\chi_{\rho}(\sigma)]^2 \, ,
\end{equation}
the multiplicity of each irrep is
\begin{equation} \label{eq:sectorDims}
    m_{\rho} = \frac{1}{e_{\rho}|H|}\sum_{\sigma \in H} \chi_{\rho}(\sigma) \tr P_{\sigma} \ ,
\end{equation}
and the projectors onto the sectors of \hyperref[eq:isotypicDecomposition]{Eq.~(\ref*{eq:isotypicDecomposition})} are
\begin{equation} \label{eq:sectorProjector}
    \Pi_\rho = \frac{d_\rho}{e_{\rho}|H|}\sum_{\sigma \in H} \chi_\rho(\sigma)\, P_\sigma \ .
\end{equation}
The trivial irrep has $\chi_0(\sigma) = 1$ for every $\sigma$, which recovers the orbit average of~\eqref{eq:Pi0} and $m_0 = r$.

These constructions apply to any real representation of $H$. For $H \subseteq \Aut(G,M')$, the \textit{accessible sectors} $\mathcal{U}_\rho \equiv \Pi_\rho(T_{\eq}M') = U_\rho \cap T_{\eq}M'$ decompose the tangent space of $M'$:
\begin{equation} \label{eq:accessibleSplit}
    T_{\eq}M' = \bigoplus_\rho \mathcal{U}_\rho \ .
\end{equation}
In particular, $\mathcal{U}_0 = T_{\eq} M$, and the remaining accessible sectors span the normal space $N_{\eq}M$ of $M$ in $M'$, as in \hyperref[eq:Ksectors]{Eq.~(\ref*{eq:Ksectors})}. In control coordinates, $H$ is represented by matrices $S_{\sigma}$, and the multiplicities and projectors of the accessible sectors can be computed from~\eqref{eq:sectorDims} and~\eqref{eq:sectorProjector} with $S_{\sigma}$ in place of $P_{\sigma}$. 

\subsection{Sector decomposition of the second variation and normal frames \label{app:comparison}}
Under the assumptions of this appendix, $P_{\sigma}$ is an isometry of $M'$ for each $\sigma \in H$, and the minimum-work protocol $\eq^\star_M \in \mathcal{P}_M$ satisfies $P_{\sigma}\eq^\star_M(t) = \eq^\star_M(t)$ and $P_{\sigma} \dot{\eq}^\star_M(t) = \dot{\eq}^\star_M(t)$. It then follows directly from the invariance of the Levi-Civita connection and the curvature tensor under isometries~\eqref{eq:connectionCommutation} that
\begin{subequations}\label{eq:permComm}
    \begin{align} 
        D'_t (P_{\sigma}X) &= P_{\sigma }D'_t X \, , \\ 
        R'(P_{\sigma}X, \dot{\eq}^\star_M )\, \dot{\eq}^\star_M  &= P_{\sigma} \, R'(X, \dot{\eq}^\star_M )\,\dot{\eq}^\star_M  \, 
    \end{align}
\end{subequations}
with $D_t'$ the covariant derivative of $M'$ along $\eq^\star_M$, $R'$ the curvature tensor of $M'$, and $X(t)$ any vector field along $\eq^\star_M$ tangent to $M'$. Since the sector projectors~\eqref{eq:sectorProjector} are linear combinations of the permutation matrices, \hyperref[eq:permComm]{Eq.~(\ref*{eq:permComm})} holds with $\Pi_{\rho}$ in place of $P_{\sigma}$. The operators therefore decouple along the sectors, from which it immediately follows that the index form also decouples:
\begin{equation} \label{eq:Idecoup}
    I(X,X)=\sum_{\rho} I(X_{\rho}, X_{\rho})
\end{equation}
with $X = \sum_{\rho}X_{\rho}$, $X_{\rho} \equiv \Pi_{\rho} X \in \mathcal{U}_{\rho}$. The tangent-normal splitting of the index form [\hyperref[eq:Isplit]{Eq.~(\ref*{eq:Isplit})}] is obtained from this with $X^{\parallel} = X_0$ and $X^{\perp} = \sum_{\rho \neq 0} X_{\rho}$. 

The Jacobi equation~\eqref{eq:jacobi} similarly splits into independent equations
\begin{equation} \label{eq:sectorJacobi}
    {D_t'}^2X_{\rho} + R'(X_{\rho}, \dot{\eq}^\star_M) \dot{\eq}^\star_M = 0\, .
\end{equation}
If there exists a nontrivial solution $X_{\rho}(t)$ with $X_{\rho}(0) = X_{\rho}(t_0) = 0$ for $t_0 < 1$, then the field $X = X_{\rho}$ (setting $X_{\rho'} =0$ for all $\rho' \neq \rho$) is a solution of the full Jacobi equation~\eqref{eq:jacobi} and $\eq^\star_M(t_0)$ is conjugate to $\eq^{\star}_M(0)$. 

The respective decompositions~\eqref{eq:R2index} and~\eqref{eq:jacobiOsc} of the second variation and the Jacobi equation are obtained by expanding each $X_{\rho}$ in a frame $\curly{\bm{n}_I^{\rho}(t)}$ of its accessible sector,
\begin{subequations} \label{eq:frame}
    \begin{align}
        \Span\curly{\bm{n}_I^{\rho}(t)}_I &= \mathcal{U}_{\rho}(t), \label{eq:fComp} \\
        \quad {\bm{n}_I^{\rho\,}}\t L^+(\eq^\star_M)\,\bm{n}^{\rho'}_J &= \delta_{\rho\rho'}\delta_{IJ}, \label{eq:fOrth} \\
        \quad D'_t \bm{n}^{\rho}_I &= 0 \ , \label{eq:fPar}
    \end{align}
\end{subequations}
which is complete~\eqref{eq:fComp}, $L^+$-orthonormal~\eqref{eq:fOrth}, and parallel along $\eq^\star_M$~\eqref{eq:fPar}. Each $\bm{n}_I^{\rho}(t)$ is a vector in $\mathbb{R}^{n+1}$ corresponding to a perturbation of the equilibrium state probabilities, and the indices $I,J$ run over the $\dim \, \mathcal{U}_{\rho} = d_{\rho}m_{\rho}$ frame vectors of sector $\rho$. Note that the sector $\mathcal{U}_{\rho}(t)$ depends on $t$ only through the tangent space $T_{\eq^\star_M(t)} M'$; the projectors $\Pi_{\rho}$ are constant. Such a frame is obtained by choosing an $L^+$-orthonormal basis of each $\mathcal{U}_{\rho}$ at $t=0$ and parallel transporting it along $\eq^\star_M$ (\hyperref[app:numerics]{App.~\ref*{app:numerics}}), which preserves orthonormality and, by \hyperref[eq:permComm]{Eq.~(\ref*{eq:permComm})}, keeps each vector within its sector.

Any variation field $X_{\rho}$ in sector $\rho$ can be written as $X_{\rho}(t) = x^I_{\rho}(t) \bm{n}^{\rho}(t)$ with the vector of coefficients $\bm{x}_{\rho}(t) \in \mathbb{R}^{d_{\rho}m_{\rho}}$. Because the frame~\eqref{eq:frame} is parallel, $D'_t X_{\rho} = \dot{x}^I_{\rho}\bm{n}^{\rho}_I$; and by the orthonormality of the frame, $\| D_t' X_{\rho}\|^2 = |\dot{\bm{x}}_{\rho}|^2$. The curvature term of the index form is therefore
\begin{subequations}
    \begin{align}
        R'(X_{\rho}, \dot{\eq}^\star_M, &\dot{\eq}^\star_M, X_{\rho}) \nonumber \\
        &= x^I_{\rho}x^J_{\rho}\, R'(\bm{n}^\rho_I, \dot{\eq}^\star_M, \dot{\eq}^\star_M, \bm{n}^{\rho}_J) \\
        &= \Wex^\star_M\,\bm{x}_{\rho}\t \, \kcur_{\rho}(t) \, \bm{x}_{\rho} \ . 
    \end{align}
\end{subequations}
Here $\kcur_{\rho}(t)\in \mathbb{R}^{(d_{\rho}m_{\rho})\times (d_{\rho}m_{\rho})}$ is the sector curvature operator of~\eqref{eq:Krho}, with normalization $\| \dot{\eq}^\star_M\|^2 = \Wex^\star_M$. The decomposition of the second variation given in~\hyperref[eq:R2index]{Eq.~(\ref*{eq:R2index})} follows immediately. Substituting the same expansion into the sector Jacobi equation~\eqref{eq:sectorJacobi} gives
\begin{equation}
    \ddot{x}^I_{\rho} \bm{n}^{\rho}_I + x^J_{\rho} \, R'(\bm{n}^{\rho}_J, \dot{\eq}^\star_M) \, \dot{\eq}^\star_M = 0\, , 
\end{equation}
and the oscillator form~\eqref{eq:jacobiOsc} of the main text is obtained by taking the $L^+$ inner product of this with $\bm{n}^{\rho}_I$. 

The thermodynamic comparison theorem~\eqref{eq:R2thresholds} requires only the eigenvalues of $\kcur_{\rho}(t)$, which are independent of the basis chosen for each $\mathcal{U}_{\rho}$. When directly integrating the Jacobi equation, it is convenient to choose the frame at $t\!=\!0$ to be adapted to the irreps of $H$ so that every operator on $\mathcal{U}_{\rho}$ that commutes with every $P_{\sigma}$ takes the block-form of \hyperref[eq:equivariantForm]{Eq.~(\ref*{eq:equivariantForm})}. Given any frame satisfying~\hyperref[eq:frame]{Eq.~(\ref*{eq:frame})}, an adapted frame may be obtained by a change of basis that simultaneously block-diagonalizes the matrices representing the action of $H$ on $\mathcal{U}_{\rho}$, for which there are standard algorithms~\cite{murotaNumericalAlgorithmBlockdiagonal2010a, maeharaNumericalAlgorithmBlockdiagonal2010}. No adapted basis was needed for the computations of \hyperref[ssec:fourspin]{Sec.~\ref*{ssec:fourspin}}, since every sector there has $e_{\rho} = 1$ and either $m_{\rho}\!=\!1$ or $d_{\rho}\!=\!1$, in which case the block form~\eqref{eq:equivariantForm} holds in any orthonormal frame. For $m_{\rho}=1$, $A_{\rho}$ is a single number $a_{\rho}$, so $I_{d_{\rho}}\otimes A_{\rho} = a_{\rho}\, I_{d_{\rho}}$ is a multiple of the identity, and for $d_{\rho} =1$ each copy is one-dimensional, so $B_{\rho} = I_1 \otimes A_{\rho} = A_{\rho}$ has no further structure.

\section{Computational methods \label{app:numerics}}
Code reproducing the computations and figures of \hyperref[ssec:fourspin]{Sec.~\ref*{ssec:fourspin}} is available at Ref.~\cite{code_ref}. Automorphism groups are computed with SageMath~\cite{sagemath}, and stabilizers, orbits, conjugacy classes, and character tables are computed with GAP~\cite{GAP4}. All differential-geometric quantities are obtained by forward-mode automatic differentiation of the metric with JAX~\cite{jax2018github}. The minimum-work protocols of the four-spin system are from Ref.~\cite{sawchuk_global_2026}, and are available at Ref.~\cite{sawchuk_2025_15678284} as arrays of state energies $\bm{V}(t_a)$ on a uniform grid of 251 times $t_a \in [0,1]$. These are converted to field coordinates $\bm{h}(t_a)$, and all computations are performed in these coordinates.  

The field control set $M_{h_i}$ is represented by
\begin{equation} \label{eq:chart}
    \eq(\bm{h}) \propto \exp\{\bm{\theta}_0 + B\bm{h}\} \ .
\end{equation}
Here $B\in \mathbb{R}^{16 \times 4}$ has elements $B_{\mu i} = s_i(\mu) = \pm 1$ equaling the value of spin $i$ in state $\mu$, and $\bm{\theta}_0\in \mathbb{R}^{16}$ is the vector of fixed coupling energies, with $(\bm{\theta}_0)_{\mu}=J_{\rm NN} \sum_{\angbr{i,j}}s_i(\mu) s_j(\mu)$ taken over nearest neighbors $\angbr{i,j}$. The Jacobian is $\mathbb{J} = \partial \eq /\partial \bm{h} = (\Deq -\eq\eq\t)B$, and the symmetric subsets $M_h$, $M_{\parallel}$, and $M_{\times}$ are obtained by restrictions of $\bm{h}$ to the appropriate linear subspaces of $\mathbb{R}^4$. $\Aut(G,M_{h_i})$ consists of the $\sigma\!\in\!\Aut(G)$ for which $P_{\sigma} \eq (\bm{h}) = \eq(\tilde{\bm{h}})$ for some $\tilde{\bm{h}}$. Permutations commute with the entrywise exponential, so $P_{\sigma} \eq(\bm{h}) \propto \exp \curly{P_{\sigma}\bm{\theta}_0 + P_{\sigma}B \bm{h}}$. The coupling energies and fields are coefficients of different-order terms in the polynomial~\eqref{eq:fourspinV} so they must be separately invariant, and therefore $\sigma$ must satisfy $P_{\sigma}\bm{\theta}_0 = \bm{\theta}_0$ and map field energies onto field energies, $P_{\sigma} B \bm{h} = B\tilde{\bm{h}}$. The action $\bm{h} \mapsto S_{\sigma}\bm{h}$ of $\sigma$ on the fields is then defined by
\begin{equation} \label{eq:inducedAction}
    P_\sigma B = B S_\sigma  \implies S_\sigma = B^+ P_\sigma B  \ .
\end{equation}
The matrices $S_{\sigma} \in \mathbb{R}^{4\times 4}$ are signed permutations of the four fields (permutations combined with spin flips), and $\Aut(G, M_{h_i}) = D_4 \times C_2$. The orthogonal projectors from the field space $\mathbb{R}^4$ to the sectors are obtained from \hyperref[eq:sectorProjector]{Eq.~(\ref*{eq:sectorProjector})}, and multiplicities $m_{\rho}$ from \hyperref[eq:sectorDims]{Eq.~(\ref*{eq:sectorDims})}, both with $S_{\sigma}$ in place of $P_{\sigma}$. 

In field coordinates, each frame vector of~\eqref{eq:frame} is $\mathbb{J}\bm{n}$ for a vector $\bm{n}\in \mathbb{R}^4$ of field perturbations, and the $L^+$-orthonormality of the frame vectors becomes orthonormality with respect to the inherited metric $\zeta = \mathbb{J}\t L^+ \mathbb{J}$ on $M_{h_i}$~\eqref{eq:pullback}. The frame at $t\!=\!0$ is computed from $\bm{N}^{\rho}(0) = Q_{\rho}(Q_{\rho}\t \zeta\, Q_{\rho})^{-1/2}$, a matrix whose columns are the frame vectors $\bm{n}^{\rho}_I(0)$ of sector $\rho$. The columns of $Q_{\rho}$ are an orthonormal basis of the sector, obtained by orthonormalizing the projections $\Pi_{\rho} \hat{e}_i$ of the single-field perturbations. The frame is transported along the protocol by fourth-order Runge--Kutta integration of the parallel-transport equation $\dot{n}^i = -\Gamma^i{}_{jk}(\bm{h}(t))\, \dot{h}^j \, n^k$, with the Christoffel symbols $\Gamma^i{}_{jk}$ obtained by automatic differentiation of the metric. We re-orthonormalize the frames with respect to $\zeta$ at each point to correct for integration error. The sector curvature operators~\eqref{eq:Krho} are then evaluated in the transported frames along the protocols. 

The sector Jacobi equation~\eqref{eq:jacobiOsc} is linear, so every solution with $\bm{x}_\rho(0) = 0$ is determined by its initial velocity. Let $\mathbbm{x}_{\rho}(t)$ be the matrix whose $j$th column is the solution with $\dot{\bm{x}}_{\rho}(0) = \hat{e}_j$. This solves the matrix ODE 
\begin{equation}\label{eq:jacMatrix}
    \begin{dcases}
        \ddot{\mathbbm{x}}_\rho + \Wex^\star_M \kcur_\rho(t) \mathbbm{x}_\rho = 0 \ ,  \\ 
        \mathbbm{x}_\rho(0) = 0, \ \ \dot{\mathbbm{x}}_\rho(0)=I \ ,
    \end{dcases}
\end{equation}
and conjugate points occur at the times $t_0 \in (0,1]$ at which $\det \mathbbm{x}_{\rho}(t_0) = 0$. With $\kcur_{\rho}(t)$ interpolated between grid points by a cubic spline, we integrate \hyperref[eq:jacMatrix]{Eq.~(\ref*{eq:jacMatrix})} as a first-order system in $(\mathbbm{x}_{\rho},\dot{\mathbbm{x}}_{\rho})$ by fourth-order Runge--Kutta on a uniform grid of 5000 steps. The columns of $(\mathbbm{x}_{\rho},\dot{\mathbbm{x}}_{\rho})$ are periodically renormalized to keep $\mathbbm{x}_{\rho}$ well conditioned, with the rescalings of $\det (\mathbbm{x}_{\rho})$ tracked and inverted on the final data. Conjugate points are identified as sign changes of $\det \mathbbm{x}_{\rho}$, which is insensitive to rescaling of $\mathbbm{x}_{\rho}$. On sectors with $m_{\rho}\!=\!1$ and $d_{\rho}\!>\!1$, $\kcur_{\rho}$ is a multiple of the identity, so the scalar equation with $\kcur_{\rho}\to \tr (\kcur_{\rho})/d_{\rho}$ is integrated instead. Sectors with both $d_{\rho} > 1$ and $m_{\rho} >1$ do not occur here and are not directly treated by the provided code.

\section{Incompleteness of the thermodynamic manifold \label{app:global}}
Here we show that the thermodynamic manifold is incomplete for common rate laws. Completeness is the standing hypothesis of most standard global theorems of Riemannian geometry~\cite{berger_panoramic_2003}, including the Hopf--Rinow theorem guaranteeing the existence of minimizing geodesics~\cite{cheeger_comparison_1975}. Its failure means that $\Wex$ need not attain its infimum on a protocol space, and prohibits the use of many global tools from comparison geometry and Morse theory without further restrictions.

Formally, a metric space is \textit{complete} if every Cauchy sequence in the space converges to a point within the space. A sequence of distributions can fail to converge in the open simplex $\simp$ only by approaching the excluded boundary, so the thermodynamic manifold is incomplete when some facet $F_{\mu} = \curly{\eq \,:\,\pi_\mu=0}$ lies at finite distance, i.e., $d(\eq, F_{\mu}) < \infty$ with 
\begin{equation} \label{eq:facetDistance}
    d(\eq, F_\mu) \, \equiv \, \inf_{\tilde{\eq}} \curly{\, \int_0^1 \md t \, \| \dot{\tilde{\eq}}\|   :  \tilde{\eq}(0)=\eq , \, \lim_{t\to1^-}\tilde{\pi}_\mu(t)=0  } \ .
\end{equation}
Denote by $\cond_\mu \equiv \sum_{\nu\sim\mu}\cond_{\mu\nu}$ the total conductance of the transitions involving state $\mu$, where $\nu \sim \mu$ runs over the neighbors of $\mu$. We will show that the distance to the facet is determined by how fast $\cond_\mu$ vanishes as $\pi_\mu \to 0$. Since any two interior points are at finite distance, it suffices to consider $\eq$ close to the facet. Assume that the conductance of the transitions involving state $\mu$ vanishes as a power $\alpha$ of $\pi_{\mu}$ while the neighbors of $\mu$ remain populated:
\begin{equation} \label{eq:uniformScaling}
    b_1\, \pi_{\mu}^{\alpha}  \leq  \cond_\mu(\eq)  \leq  b_2\, \pi_{\mu}^{\alpha}
    \quad (\eq : \pi_\mu \ll 1) \ .
\end{equation}
This is assumed to hold only when the probabilities of the neighbors of $\mu$ are bounded away from zero, and in general $b_1, b_2$ depend on that bound. For Glauber rates, $\cond_{\mu\nu} = \pi_\mu\pi_\nu /(\pi_\mu + \pi_\nu) \to \pi_\mu$ as the state empties, so $\alpha\!=\!1$. For square-root rates $w_{\mu\nu} \propto \sqrt{\pi_\mu /\pi_\nu}$~\cite{lieSquareRootApproximation2013, dixit_inferring_2015}, $\cond_{\mu\nu}\propto \sqrt{\pi_\mu\pi_\nu}$, and so $\alpha\!=\!1/2$. Generally, a rate that vanishes as $w_{\mu\nu} \propto (\pi_{\mu}/\pi_{\nu})^{\alpha}$ gives $\cond_{\mu\nu} \propto \pi_{\mu}^{\alpha}\pi_{\nu}^{1-\alpha}$. 

The length $\ell[\tilde{\eq}]$ of any protocol that empties the state provides an upper bound on $d(\eq, F_{\mu})$. Take $\pi_\mu\!=\!a$ sufficiently small so that~\eqref{eq:uniformScaling} applies, and drain state $\mu$ into its neighbors in proportion to the conductances of the edges joining them, $\dot{\tilde{\pi}}_{\nu} = \cond_{\mu\nu} / \cond_\mu$. By construction, $\dot{\tilde{\pi}}_{\nu}$ is $-1$ for $\nu\!=\!\mu$ and is zero if $\nu$ is not a neighbor of $\mu$. The conjugate potential is directly read off as $\phi_\nu = - \delta_{\mu\nu}/\cond_{\mu}$ from the definition $\dot{\eq} = L \phi$, so $\|\dot{\tilde{\eq}}\|^2 = \sum_\nu \phi_\nu \dot{\tilde{\pi}}_\nu = 1 /\cond_{\mu}$ and thus 
\begin{equation} \label{eq:facetUpper}
    d(\eq, F_{\mu}) \leq \ell[\tilde{\eq}] = \int_0^a \frac{\md \pi_{\mu}}{\sqrt{\cond_{\mu}}} \leq \frac{1}{\sqrt{b_1}}\int_0^a \frac{\md \pi_{\mu}}{\pi_\mu^{\alpha /2}} \ .
\end{equation}
For a lower bound of~\eqref{eq:facetDistance}, we must also consider protocols that empty multiple states simultaneously. Let $S$ be the set of all states that a protocol empties, so that the states $\Omega \setminus S$ remain populated. Denote by $\pi_S \equiv \sum_{\nu\in S}\pi_\nu$ the total probability in $S$ and by $\cond_S \equiv \sum_{\mu \in S}\sum_{\nu \notin S} \cond_{\mu\nu}$ the total conductance of the edges leaving $S$. With potential $\bm{\phi} = L^+\dot{\eq}$, write $\dot{\pi}_\mu = \sum_{\nu\sim \mu} \cond_{\mu\nu}(\phi_\mu - \phi_\nu)$. Summing over the states in $S$, any edge joining two states in $S$ appears twice with opposite signs and cancels, leaving
\begin{subequations} \label{eq:speedBound}
    \begin{align}
        |\dot{\pi}_S|^2 &= \bigg|\sum_{\mu\in S}\sum_{\nu \notin S} \cond_{\mu\nu} (\phi_\mu - \phi_\nu )\bigg|^2 \\
        & \leq \cond_S\sum_{\mu\in S}\sum_{\nu \notin S} \cond_{\mu\nu}(\phi_\mu - \phi_\nu)^2 \\
        &\leq \cond_S\sum_{(\mu,\nu)\in \mathcal{E}} \cond_{\mu\nu} (\phi_\mu - \phi_\nu)^2 \\
        &= \cond_S \| \dot{\eq} \|^2 \, . 
    \end{align}
\end{subequations}
The first inequality is Cauchy--Schwarz, and the second follows from nonnegativity of the summand. Assume first that the exponent $\alpha$ in~\eqref{eq:uniformScaling} is the same for every state, which holds, e.g., if every transition in the network obeys the same rate law. Each edge leaving $S$ joins an emptying state $\mu$ to a populated state, and since each conductance depends only on the probabilities of the two states it joins, \hyperref[eq:uniformScaling]{Eq.~(\ref*{eq:uniformScaling})} bounds each of these conductances separately, $\cond_{\mu\nu}\leq b_2 \pi^{\alpha}_\mu \leq b_2 \pi^\alpha_S$. There are at most $|\mathcal{E}|$ such edges, so 
\begin{equation} \label{eq:condBound}
    \cond_S \leq |\mathcal{E}|\,b_2\,\pi^\alpha_S \, .
\end{equation}
A protocol $\tilde{\eq}$ that empties $S$ must pass through all probabilities $\pi_S \in (0,a)$ [again, $a$ is assumed small enough that~\eqref{eq:uniformScaling} holds]. Then the length of any such protocol satisfies
\begin{equation} \label{eq:setLower}
    \ell[\tilde{\eq}] \geq \int \md t \, \frac{|\dot{\tilde{\pi}}_S |}{\sqrt{\cond_S}} \geq \frac{1}{\sqrt{b_2|\mathcal{E}|}}\int_0^a \frac{\md \pi_S}{\pi_S^{\alpha/2}} \ .
\end{equation}
The first inequality is~\eqref{eq:speedBound} and the second is~\eqref{eq:condBound}. Any protocol that reaches $F_{\mu}$ empties a set $S$ containing $\mu$, so~\eqref{eq:setLower} lower bounds the distance to $F_{\mu}$:
\begin{equation} \label{eq:facetLower}
    d(\eq, F_{\mu}) \geq \frac{1}{\sqrt{b_2|\mathcal{E}|}}\int_0^a \frac{\md \pi_S}{\pi_S^{\alpha/2}} \ .
\end{equation}
Together,~\eqref{eq:facetLower} and~\eqref{eq:facetUpper} bound the distance to the facet by the same integral. The facet therefore lies at infinite distance if and only if $\alpha \geq 2$. For Glauber rates ($\alpha\!=\!1$) and square-root rates ($\alpha\!=\!1/2$), the integral converges and the thermodynamic manifold is incomplete. If the exponent $\alpha$ in~\eqref{eq:uniformScaling} differs between states, the upper bound~\eqref{eq:facetUpper} holds with $\alpha$ the exponent of state $\mu$, and the lower bound~\eqref{eq:facetLower} holds with $\alpha$ the smallest such exponent among states in $S$. The thermodynamic manifold is therefore complete if and only if every exponent is at least two.

We demonstrate this explicitly for a two-state system. With $\eq(p) = (p,1-p)\t$ and one edge of conductance $\cond(p)$, the friction metric is $\md s^2 = \md p^2 / \cond(p)$. For Glauber rates, the conductance is $\cond(p)= p(1-p)$ and the distance between $\eq(p)$ and $\eq(q)$ for arbitrary $p < q \in (0,1)$ is 
\begin{equation}\label{eq:glaub2dist}
    d(p,q)= \arcsin(2q-1) - \arcsin(2p-1) \ .
\end{equation}
In the limits $p\!\to\!0$ and $q\!\to\!1$, this gives $d(0,1) = \uppi$, so the manifold is incomplete. This result may have been anticipated: The friction tensor for a two-state system under Glauber dynamics is just the Fisher--Rao metric, so the thermodynamic manifold has the geometry of a quarter circle of radius $2$ in the Euclidean plane (\hyperref[ssec:RaoIllust]{Sec.~\ref*{ssec:RaoIllust}}), hence total arc length $\uppi$.

%

\end{document}